\documentclass[useAMS,usenatbib]{mn2e}
\usepackage{amstext}
\usepackage{amssymb}
\usepackage{amsmath, float}
\usepackage{graphicx}
\usepackage{txfonts}
\usepackage{amsmath,mathtools,revsymb,amssymb}
\usepackage{booktabs,multirow,calc,xspace,rotating}

\newcommand       \be           {\begin{equation}}
\newcommand       \ee           {\end{equation}}
\newcommand       \bea          {\begin{eqnarray}}
\newcommand       \eea          {\end{eqnarray}}

\usepackage[T1]{fontenc}
\usepackage[english]{babel}
\usepackage{lmodern}
\usepackage{microtype}
\usepackage{mathtools}
\usepackage{cleveref}
\usepackage[utf8]{inputenc}

\usepackage{bm}
\let\vec\bm
\newcommand{\tvec}[1]{\tilde{\vec{#1}}}

\usepackage{mleftright}
\let\left\mleft
\let\right\mright

\usepackage{color}
\usepackage{comment}
\includecomment{comment}
\newcommand{\notes}[1]{}
\newcommand{\pder}[2]{\frac{\partial{#1}}{\partial{#2}}}

\usepackage{xspace}

\newcommand{\sums}{\sum\nolimits_s}

\newcommand{\va}{v_a}

\newcommand{\vt}{v_t}

\newcommand{\omc}{\omega_c}
\newcommand{\omcs}{\omega_{cs}}

\newcommand{\omg}{\omega_g}

\newcommand{\ex}{\vec{e}_x}
\newcommand{\ey}{\vec{e}_y}
\newcommand{\ez}{\vec{e}_z}

\begin{document}

\title[Weak Field Instabilities in Differential Rotating Plasmas]
{Linear Instabilities Driven by Differential Rotation in Very Weakly Magnetized Plasmas}
\author[E. Quataert, T. Heinemann, \& A. Spitkovsky]{E. Quataert$^{1}$\thanks{E-mail: eliot@berkeley.edu}, T. Heinemann$^{2}$, \& A. Spitkovsky$^{3}$   \\ $^{1}$ Astronomy Department and Theoretical Astrophysics  Center, University of California, Berkeley \\ $^{2}$ Kavli Institute for Theoretical Physics, University of California, Santa Barbara \\ $^{3}$ Department of Astrophysical Sciences, Princeton University}

\maketitle

\begin{abstract}

We study the linear stability of weakly magnetized  differentially rotating plasmas { in both collisionless kinetic theory and Braginskii's theory of collisional, magnetized plasmas.}   We focus on the very weakly magnetized limit in which $\beta \gtrsim \omega_c/\Omega$, where $\beta$ is the ratio of thermal to magnetic energy and $\omega_c/\Omega$ is the ratio of the cyclotron frequency to rotation frequency.   This regime is important for understanding  how astrophysical magnetic fields originate and are amplified at high redshift.   We show that the single instability of fluid theory - the magnetorotational instability mediated by magnetic tension -  is replaced by two distinct instabilities, one associated with  ions and one with electrons.  Each of these has a different way of tapping into the free energy of differential rotation.   The ion instability is driven by viscous transport of momentum across magnetic field lines due to a finite ion cyclotron frequency (gyroviscosity); {  the fastest growing modes have wavelengths significantly longer than MHD and Hall MHD predictions.}   The electron instability is a whistler mode driven unstable by the temperature anisotropy generated by  differential rotation; the growth time can be orders of magnitude shorter than the rotation period.  The electron instability is an example of a broader class of instabilities that tap into the free energy of differential rotation or shear via the temperature anisotropy they generate.   {  We briefly discuss the application of our results to the stability of planar shear flows and show that such flows are linearly overstable in the presence of fluid gyroviscosity.   We also briefly describe the implications of our results for magnetic field amplification in the virialized halos of high redshift galaxies.}

\end{abstract}

\section{Introduction}

\citet{Balbus1991} demonstrated that even energetically weak magnetic fields can be dynamically important:   when the magnetic energy in a plasma is very small compared to the thermal or rotational energies, magnetic tension is nonetheless  important for small wavelength fluctuations.   For the specific case of the  magnetorotational instability (MRI) in  magnetohydrodynamics (MHD), the fastest growing mode has a
 growth rate of $\sim \Omega$ (independent of the field strength) and a wavelength $\sim v_a/\Omega$, where $\Omega$ is the rotation rate and $v_a$ is the Alfv\'en speed.   

For sufficiently weak magnetic fields the fastest growing mode predicted by the MHD theory of the MRI can have a wavelength sufficiently small that the single fluid MHD approximation breaks down.    Unless there are large primordial magnetic fields generated in the early Universe, weak fields of this magnitude will inevitably be the `initial condition' during the formation of the first stars and galaxies.  A natural question is how such magnetic fields get amplified to the point where MHD becomes a plausible model of the plasma dynamics?   And are magnetic stresses  dynamically important for  the formation of even the first astrophysical objects?  {  In this paper we address aspects of this problem by considering the linear stability of differentially rotating plasmas with very weak magnetic fields in both collisionless kinetic theory  and the collisional magnetized theory of \citet{Braginskii1965}.\footnote{For brevity we shorten `collisionless kinetic theory' to `kinetic theory' in most places in this paper.   And by Braginskii's collisional magnetized theory we specifically mean the anisotropic viscous transport present in a collisional plasma when the cyclotron frequency is larger than the collision frequency (see \S \ref{sec:gyroviscous}).}  }  This allows us to begin to address how very weak initial magnetic fields can be amplified even when ions and electrons are only partially magnetized (in the sense of having Larmor radii comparable to the size of the system under study).   

There is a significant literature studying extensions of the MRI beyond the ideal MHD approximation (e.g., \citealt{Blaes1994,Wardle1999,Quataert2002,Ferraro2007}).   One approach to studying the weak field limit is based on Hall MHD, which takes into account the difference between the ion and electron dynamics when fluctuations have timescales comparable to or shorter than the ion cyclotron period.   The MRI persists even in this limit (as a destabilized whistler wave), with the same maximum growth rate as in MHD \citep{Wardle1999,Balbus2001}.   The  Hall MHD theory of the MRI is motivated primarily by the application to protostellar disks where the plasmas are collisional but deviations from MHD are due to the very low density of charge carriers (e.g., \citealt{Lesur2014}).    By contrast, Hall MHD does not provide a good description of low-collisionality weakly magnetized plasmas with $\beta \gg 1$, as \citet{Ferraro2007}  emphasized in the context of the MRI (see, e.g., \citealt{Howes2009} for a more general discussion of some of the limitations of Hall MHD).   Our approach in this paper is to carry out a kinetic linear stability calculation, valid so long as the fluctuations of interest have wavelengths smaller than the electron and proton mean free paths.   Our work draws heavily on that of  \citet{HQ2014} (hereafter HQ), who studied the linear kinetic theory of local instabilities in differentially rotating plasmas.   {  We also show that the kinetic ion instability described in this paper has a fluid analogue in which both planar shear flows and differentially rotating plasmas are destabilized by gyroviscosity.}

In \S \ref{sec:HQ} we summarize the aspects of HQ's formalism important for our analysis.  We then present numerical solutions for linear instabilities of differentially rotating plasmas for the case of weak magnetic fields aligned or anti-aligned with the rotation axis of the system (\S \ref{sec:numerics}); \S \ref{sec:analytics-i} \& \ref{sec:analytics-e} present analytical approximations to these numerical instability calculations and elucidate the physics.   {   In \S \ref{sec:shear} we show how the results derived in \S \ref{sec:results} can be applied to the problem of planar shear flows in addition to differentially rotating plasmas.   In \S \ref{sec:halo} we briefly describe the application of our results to magnetic field amplification in the virialized halos of high redshift galaxies.}  Finally, in \S \ref{sec:discussion} we summarize and discuss our results.

\section{The Linear Theory of the Shearing Sheet in a Collisionless Plasma}
\label{sec:HQ}

HQ derived the linear theory of the shearing sheet for a collisionless plasma.   We review here some of their results that are important for our analysis but we largely defer to their paper for details.   For consistency, we utilize the same notation as HQ throughout (including their use of SI units for electromagnetism).   We use the subscript $s$ to represent a particular particle species (e.g., electron, ion) but drop the subscript for clarity when it is not required.

We describe the dynamics of a local patch of a differentially rotating flow in a rotating reference frame.  The coordinate system is locally cartesian and we neglect vertical stratification.   The coordinate system is such that the rotation axis is in the $\ez$ direction so that the background rotational velocity is along $\ey$.   The equilibrium magnetic field is given by $\vec{B} = B_y \ey + B_z \ez$.    The equilibrium thus corresponds to a uniform density $n$ in $x, y,$ and $z$, with the bulk motion of each plasma species as viewed in the rotating reference frame given by
\begin{equation}
  \label{eq:shear-flow}
  \vec{u} = -q\Omega x \ey \ \ \ \ {\rm where} \ \ \ \ q \equiv -\frac{d \Omega}{d \ln r}.
\end{equation}
The Maxwellian distribution function with uniform density and a background velocity given by equation \ref{eq:shear-flow} is 
\begin{equation}
  \label{eq:maxwellian}
  f(\mathcal{K}) =
  \frac{n\exp(-\mathcal{K}/\vt^2)}{(2\pi)^{3/2}\vt^3\sqrt{1-\Delta}},
\end{equation}
where $\vt=\mathrm{const}$ is the thermal velocity,
\begin{equation}
  \label{eq:gyration-energy}
  \mathcal{K} =
  \frac{1}{2}\left[v_x^2 + \frac{(v_y + q\Omega x)^2}{1-\Delta} + v_z^2\right]
\end{equation}
is the gyration energy (the difference between a particle's energy and the energy of a hypothetical particle with the same angular momentum but on a circular orbit), 
\begin{equation}
  \label{eq:tidal-anisotropy}
  \Delta = \frac{q\Omega}{\omega_{c} b_z + 2 \Omega},
\end{equation}
is the tidal anisotropy and $\omega_{c} = e B/m$ is the cyclotron frequency.  The sign conventions used here are that $e$ can be positive or negative while $B > 0$; $b_z = B_z/B$ denotes the component of the total field along the rotation axis.   Note that $b_z = \pm 1$ for a purely vertical field, representing the field aligned ($b_z = 1$) or anti-aligned ($b_z = -1$) with respect to the rotation axis.

Equation \ref{eq:gyration-energy} shows that $\Delta$ in equation \ref{eq:tidal-anisotropy} is the species-dependent temperature anisotropy imposed by the background differential rotation.   This is the level of temperature anisotropy inevitably created by differential rotation in a collisionless plasma and is distinct from the more familiar temperature anisotropy with respect to a mean magnetic field.    The temperature anisotropy implied by equation \ref{eq:gyration-energy} is 
\be
\label{eq:temp-anisotropy}
T_x =  \frac{T_y}{1 - \Delta}.
\ee
Depending on the sign of $\Delta$, $T_y$ can be either less than or larger than $T_x$.   
For the case of a vertical magnetic field, the tidal  anisotropy is entirely in the plane {\em perpendicular} to the mean magnetic field, i.e., it is distinct from the typical temperature anisotropy considered in homogeneous magnetized plasmas.    By contrast, for the more general case of $B_z \ne 0$ and $B_y \ne 0$, the tidal  anisotropy includes both an anisotropy in the plane perpendicular to the magnetic field and anisotropy with respect to the mean magnetic field.    Finally, we note that the dynamics of differential rotation only imposes an $x-y$ ($r-\phi$) temperature anisotropy.   In equation \ref{eq:maxwellian} we have for simplicity taken $T_z = T_x$ although this in general need not be true.   Relaxing the restriction to $T_z = T_x$ would generate an even larger class of instabilities driven by temperature anisotropy than those that we present in this paper.

For an unmagnetized plasma, $\Delta = q/2$, which is also the standard anisotropy of stellar dynamics \citep{Shu1969}; this corresponds to $T_y = T_x/4$ for a point mass potential with $q = 3/2$.     In the opposite guiding center limit in which $\omega_{c} b_z/\Omega \rightarrow \infty$, $\Delta \rightarrow 0$.    In this paper, we are interested in the weak field limit, which corresponds to  finite $\Delta \ne 0$.   

There is an extensive literature on instabilities driven by temperature anisotropies in a homogeneous (non-shearing) plasma, both with and without a mean magnetic field (e.g., \citealt{Weibel1959,Gary1993}).  In what follows we shall show the  surprising connection between these instabilities and the weak field limit of the MRI in a collisionless plasma.

HQ derived the linear dispersion relation for charge-neutral axisymmetric ($k_y = 0$) perturbations about the equilibrium state described above, with perturbations $\propto \exp(-i \omega t + \vec{k} \cdot \vec{x})$.  Note that $\mathrm{Im}\,\omega > 0$ corresponds to instability.  The dispersion relation follows from linearizing the Vlasov equation and Maxwell's equations in the shearing sheet.  The calculation is carried out using a suitable choice of velocity coordinates that make it possible to draw on the extensive existing linear Vlasov theory for a uniform plasma (e.g., \citealt{Ichimaru1973}).  The resulting dispersion relation is \be
\mathbf{D}\cdot \left( \mathbf{1} -  \frac{q\Omega}{i\omega} \, \ex\ey \right) \cdot\delta\tvec{E}=0
\ee
where $\tvec{E} = \vec{E} - q\Omega x\vec{e}_y\times\vec{B}$ is the electric field as
seen by an observer that is locally at rest with respect to the background
shear flow.   Note that  $\det\left(\mathbf{1} - q \Omega/i\omega \, \ex \ey\right) = 1$ so that the dispersion relation is determined solely by the dispersion tensor, which is given by
\begin{align}
\label{eq:shearing-sheet-dispersion-relation}
 \mathbf{D} = & \
\left(k^2\mathbf{1} - \vec{k}\vec{k} - i\omega\mu_0\vec{\sigma}\right) \, \va^2 \\
- &  \ \ 2q\Omega^2\sums\frac{n_s m_s}{\rho}
\frac{\omcs \, b_z}{\omcs b_z + 2\Omega}\ey\ey \nonumber
\end{align}
where the sum is over each species $s$, $\rho=\sums{}n_sm_s$ is the total mass density, $\va^2=B^2/(\mu_0\rho)$ is the square of the Alfv\'en speed, and $\vec{\sigma}$ is the linear shearing sheet conductivity tensor (summed over species) calculated by HQ.   The latter is related to the linear response tensor of each species $\mathbf{\Lambda}_s$ using
\begin{equation}
 \label{eq:conductivity-sheet}
\vec{\sigma} = -\frac{1}{i\omega}\sums\frac{e_s^2n_s}{m_s}
 \mathbf{Q}_s\cdot\mathbf{\Lambda}_s\cdot\mathbf{Q}_s.
\end{equation}
where  \begin{equation}
\label{eq:aniso-tensor}
\mathbf{Q}_s = \ex\ex + \ey\ey\sqrt{1-\Delta_s} + \ez\ez.
\end{equation}
The full plasma response tensor $\mathbf{\Lambda}_s$ for each species for the distribution function in equation \ref{eq:maxwellian} is given in HQ's eqs 49-53.   This in turn depends on the plasma response function $W(\xi)$  \citep{Ichimaru1973}, which is related to the standard plasma dispersion function $Z$ via
$W(\zeta)=1+\xi{}Z(\xi)$ with $\zeta=\sqrt{2}\,\xi$. 

\section{Linear Theory with Kinetic Ions and Electrons:   {$\MakeLowercase{\vec{k}} \parallel \vec{\Omega} \parallel \vec{B}$}}
\label{sec:results}

HQ numerically solved for the linear theory of the MRI for {$\vec{k} \parallel \vec{\Omega} \parallel \vec{B}$} (including the case of $\vec{B}$ and $\vec{\Omega}$ anti-parallel) for the case of kinetic ions and cold, massless electrons.   Here we generalize their results and show that  in the weak field limit  accounting for kinetic electrons substantially changes the physics and growth rates.   We also clarify some of the physics of modes driven by kinetic ions with finite cyclotron frequency, a case  considered by \citet{Ferraro2007} \& HQ.    We take $q = 3/2$ in all of our numerical solutions.   Our method of numerically solving for the `MRI' branch of the full kinetic dispersion is described in \S 5.3.3 of HQ.  One key approximation is that we restrict ourselves to searching for purely growing modes (i.e., instability not overstability).   

\subsection{Numerical Solutions}
\label{sec:numerics}

\begin{figure*}
\centering\includegraphics{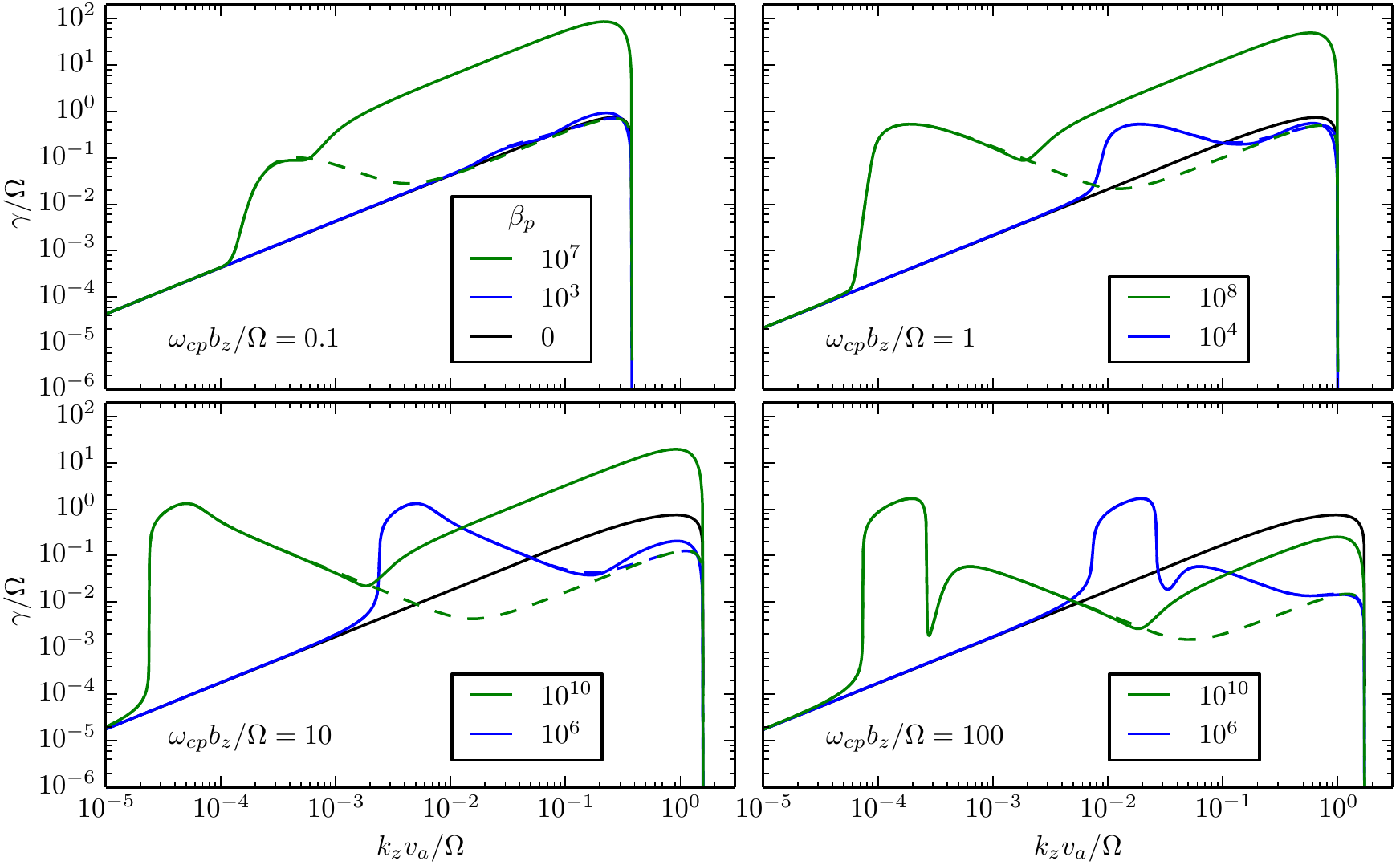}
\caption{Numerical growth rates $\gamma$ predicted by the kinetic ion and electron dispersion relation (solid colored lines) for $\vec{k} \parallel \vec{\Omega} \parallel \vec{B}$ for different values of $\beta_p = \beta_e$ and $\omega_{cp}/\Omega$ (taking $T_e = T_p$ and $m_p = 1836 \, m_e$).   Also shown is the dispersion relation for kinetic ions but cold, massless electrons (dashed lines) and the Hall MHD (cold ion, cold massless electron) dispersion relation (black solid lines, which correspond to $\beta_p = 0$).   For weak magnetic fields (high $\beta$) kinetic ion physics generates an instability at much longer wavelengths relative to that predicted by MHD or Hall MHD.   This produces the distinctive peaks in the growth rate at low $k_z v_a/\Omega$ (see \S \ref{sec:analytics-i} and eq. \ref{eq:DR-ion} for the interpretation).   Moreover,  at high $\beta$ kinetic electrons lead to substantially enhanced growth rates $\gg \Omega$ at $k_z v_a/\Omega \sim 1$.  This is an electron whistler instability driven by the temperature anisotropy in a differentially rotating plasma (see \S \ref{sec:analytics-e} and eqs. \ref{eq:DR-par} \& \ref{eq:om-max}).\label{fig:DR}}
\end{figure*}

\begin{figure}
\centering\includegraphics{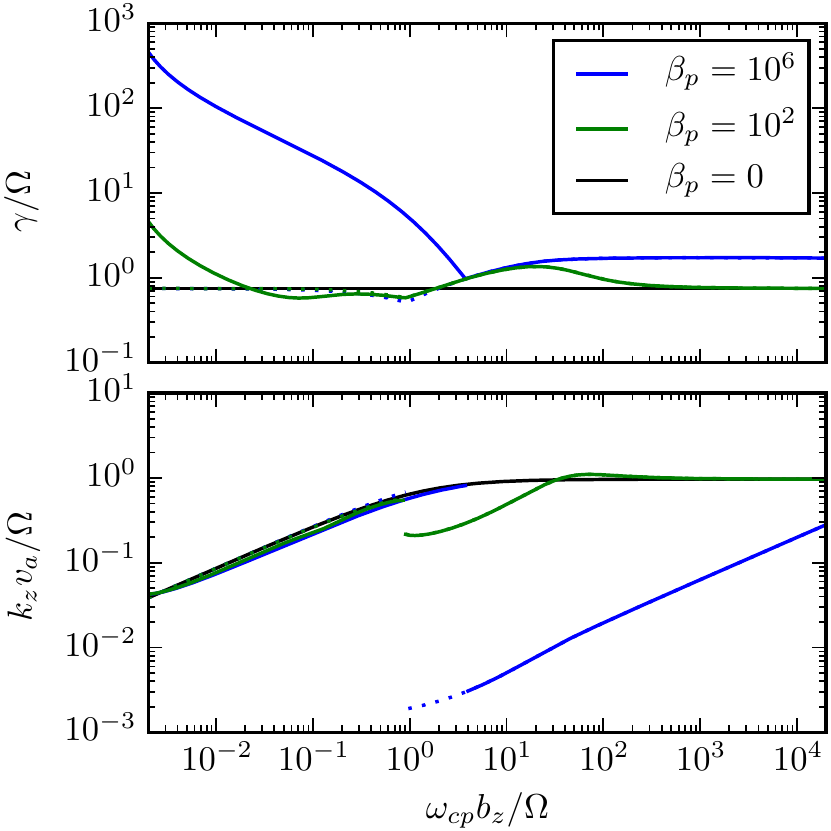}
\caption{Growth rate and wavevector of the fastest growing mode for the kinetic ion and electron dispersion relation (solid colored lines) as a function of proton cyclotron frequency $\omega_{cp}/\Omega$ for $\vec{k} \parallel \vec{\Omega} \parallel \vec{B}$, $T_e = T_p$, and $m_p = 1836 \, m_e$.    Dashed lines show solutions for kinetic ions and cold, massless electrons and black solid lines show the Hall MHD solution (the cold ion and cold, massless electron limit of kinetic theory, i.e., $\beta_p = 0$).    At high $\beta_p \gtrsim \omega_{cp}/\Omega \gtrsim 1$, kinetic ions shift
the fastest growing mode to substantially longer wavelengths than predicted by Hall MHD.   This instability is driven by ion transport of momentum (`viscosity') rather than magnetic tension (see eqs. \ref{eq:DR-ion}-\ref{eq:kmax-ion} in \S \ref{sec:analytics-i}).   In addition, at low $\omega_{cp}/\Omega$, i.e., for very weakly magnetized plasmas, the solutions with kinetic electrons differ significantly from the Hall MHD or kinetic ion solutions, with much faster growth rates.   This is the electron whistler instability driven by the temperature anisotropy in a differentially rotating plasma (see eq. \ref{eq:kmax}-\ref{eq:om-max} in \S \ref{sec:analytics-e}).\label{fig:omc}}
\end{figure}

\begin{figure}
\centering\includegraphics{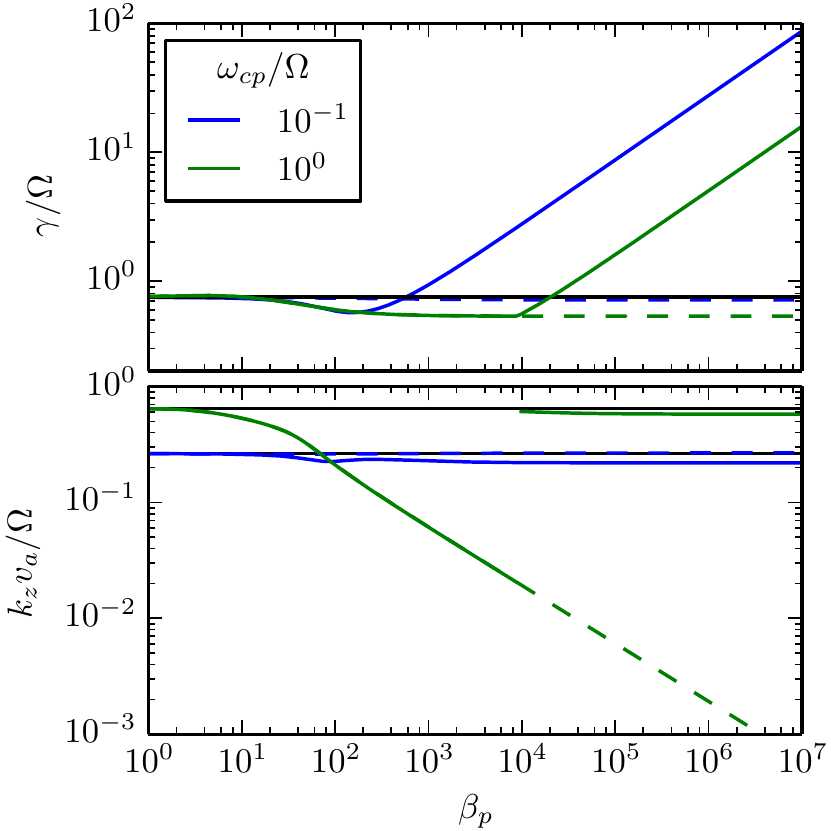}
\caption{ Growth rate and wavevector of the fastest growing mode for the kinetic ion and electron dispersion relation (solid colored lines) as a function of $\beta_p = \beta_e$ (taking $\vec{k} \parallel \vec{\Omega} \parallel \vec{B}$, $T_e = T_p$, and $m_p = 1836 \, m_e$).   Dashed lines show solutions for kinetic ions and cold, massless electrons while the black solid lines show the Hall MHD solution, which corresponds to $\beta_p = 0$.   At low $\beta_p$, Hall MHD provides a reasonable approximation to the full kinetic solution, as shown by the convergence of the kinetic and Hall MHD solutions at low $\beta_p$.  For weak fields, however, i.e., high $\beta_p$, the fastest growing modes have growth rates much larger than predicted by either Hall MHD or a kinetic ion, fluid electron theory (and with $\gamma \propto \beta^{1/2}$).   
This rapid growth is an electron whistler instability  driven by the temperature anisotropy in a differentially rotating low collisionality plasma (see eq. \ref{eq:kmax}-\ref{eq:om-max} in \S \ref{sec:analytics-e}).\label{fig:beta}}
\end{figure}

\begin{figure}
\centering\includegraphics{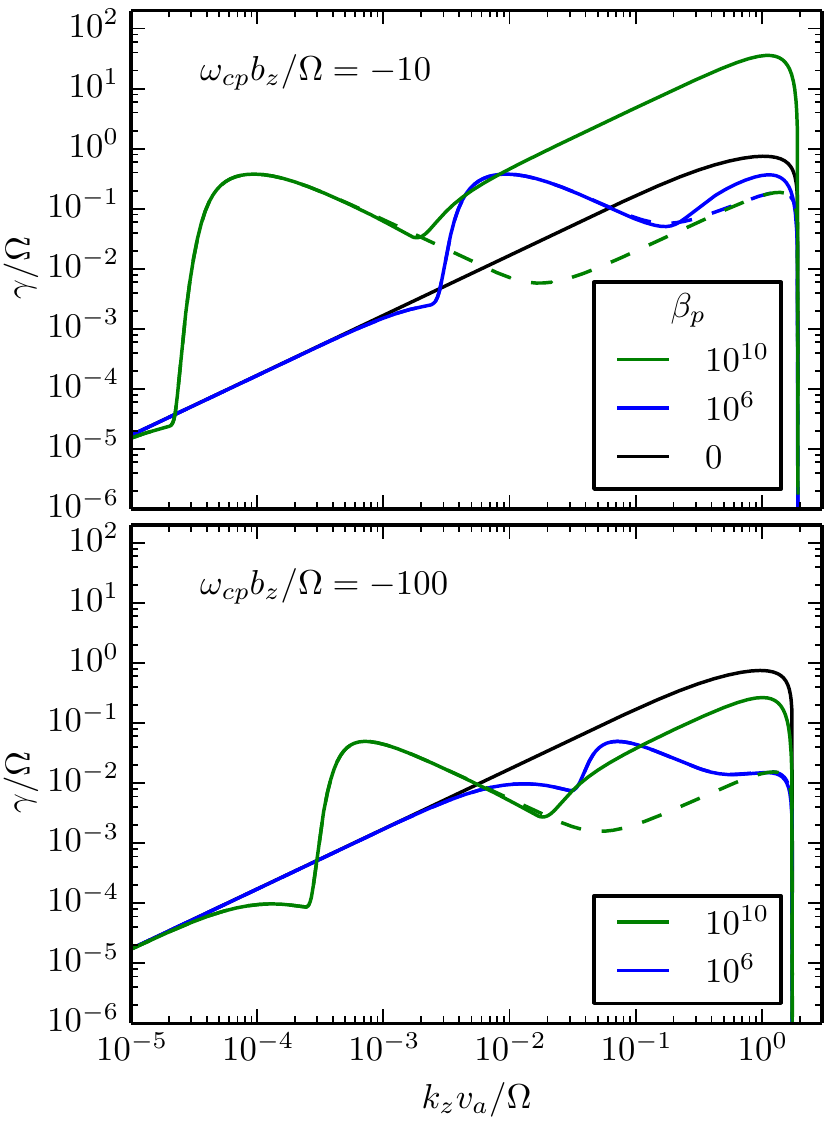}
\caption{Numerical growth rates $\gamma$ predicted by the kinetic ion and electron dispersion relation as a function of $k_z v_a/\Omega$ (solid colored lines) for $b_z = -1$, i.e., $\vec{\Omega}$ antiparallel to $\vec{B}$, taking $T_e = T_p$ and $m_p = 1836 \, m_e$.  Dashed lines show solutions for kinetic ions and cold, massless electrons while black solid lines show the Hall MHD solution.    Comparison to Figure \ref{fig:DR} shows that the kinetic ion instability 
at long wavelengths ($k_z v_a \ll \Omega$) is sensitive to the sign of $\vec{\Omega}\cdot\vec{B}$ while the kinetic electron mode at $k_z v_a \sim \Omega$ is not.   For the antiparallel case considered here, there are no growing modes for $\omega_{cp}/\Omega \lesssim 1$, unlike for the case of $\vec{\Omega} \parallel \vec{B}$ shown in Figure \ref{fig:DR} where we show solutions for $\omega_{cp}/\Omega = 0.1$ and 1.   These results are explained analytically in \S \ref{sec:analytics-i} \& \ref{sec:analytics-e}.    \label{fig:DR-anti}}
\end{figure}

Figure \ref{fig:DR} shows growth rates as a function of $k_z v_a/\Omega$ for several different values of $\omega_{cp}/\Omega$ and $\beta_p = \beta_e$.    Figures \ref{fig:omc} \& \ref{fig:beta} then show the maximum growth rate and the wavelength of the fastest growing mode as a function of $\omega_{cp}/\Omega$ and $\beta$, respectively.     In all three Figures, we take $b_z = 1$, i.e., $\vec{\Omega}$ and $\vec{B}$ are parallel.   The anti-parallel case is shown in Figure \ref{fig:DR-anti} discussed below.    Figures \ref{fig:DR}-\ref{fig:beta} also show solutions for three different approximations to the physics:   Hall MHD, i.e., the cold ion and cold, massless electron limit of kinetic theory  (black solid lines), kinetic ions and cold, massless electrons (dashed colored lines), and kinetic ions and electrons (solid colored lines).   In all of our calculations with kinetic electrons, we take $T_p = T_e$ (and hence $\beta_p = \beta_e$) and $m_p = 1836 \, m_e$. 

The Hall MHD results shown in Figures \ref{fig:DR}-\ref{fig:beta} reproduce the well-known MRI results in the literature, with the maximum growth rate of $q \Omega/2$ occurring at $k_z v_a/\Omega \sim \min(1, [\omega_{cp}/\Omega]^{1/2})$ \citep{Wardle1999,Balbus2001}.   Figures \ref{fig:DR}-\ref{fig:beta} show, however, that the physics is very different for the case of kinetic ions and cold, massless electrons.   Most notably, the maximum growth rate can reach $(-d \Omega^2/d \ln r)^{1/2} = \sqrt{3} \Omega$ (for $q = 3/2$) for $\beta \gg \omega_{cp}/\Omega \gtrsim 1$ and the maximum growth occurs at a very different wavelength, roughly $k v_a/\Omega \sim \beta^{-1/2} (\omega_{cp}/\Omega)^{1/2}$, which is equivalent to $k v_{tp}/\Omega \sim (\omega_{cp}/\Omega)^{1/2}$.  These significant differences between the kinetic ion and Hall MHD results are  particularly striking in the low $k$ peak in the dispersion relation in Figure \ref{fig:DR} and the $\omega_{cp}/\Omega \gtrsim 10$ solution for the fastest growing mode in Figure \ref{fig:omc}.   Note, moreover, that although this kinetic ion mode is {\em not} the fastest growing mode in the presence of kinetic electrons, the low $k$ peak in the dispersion relation is not significantly modified by the inclusion of kinetic electrons.   It is an essentially ion driven mode.    We elucidate the physics of this ion-driven instability in \S \ref{sec:analytics-i}.    

Finally, we turn to the case of kinetic electrons.  Figures \ref{fig:DR}-\ref{fig:beta} show that the inclusion of kinetic electrons introduces a fundamentally new unstable mode.  Remarkably, although the wavelength of the fastest growing mode is similar to the case of Hall MHD, the growth rate is far faster, with $\gamma \gg \Omega$ for high $\beta_e$ and finite $\omega_{cp}/\Omega$.  Moreover the growth rate increases $\propto \beta_e^{1/2}$ at fixed $\omega_{cp}/\Omega$ (Fig. \ref{fig:beta}).  This can exceed by orders of magnitude the previously known fastest growth rate for modes driven by the free energy in differential rotation ($-d \Omega^2/d\ln r$; \citealt{Quataert2002}).  We explain these results analytically in \S \ref{sec:analytics-e}.

Figure \ref{fig:DR-anti} shows growth rates as a function of $k_z v_a/\Omega$ for case of $b_z = -1$, i.e., $\vec{\Omega}$ and $\vec{B}$ anti-parallel.  As before, we consider several different values of $\omega_{cp}/\Omega$ and $\beta_p = \beta_e$ and show the dispersion relation curves for same three physics cases considered in Figures \ref{fig:DR}-\ref{fig:beta}, namely Hall MHD, kinetic ions and cold, massless electrons, and kinetic ions and electrons.  A comparison of Figures \ref{fig:DR} and \ref{fig:DR-anti} shows that the ion-driven instability at $k_z v_a/\Omega \ll 1$ is sensitive to the sign of $b_z$, with rapid growth only for $b_z = 1$.  By contrast, the kinetic electron instability at $k_z v_a \sim \Omega$ is insensitive to the sign of $b_z$.  We explain these results analytically in the following subsections.

For the antiparallel case shown in Figure \ref{fig:DR-anti} we were unable to find growing modes for $\omega_{cp} \lesssim \Omega$, unlike for $b_z = 1$, where there can be rapid growth associated with kinetic electrons (Fig. \ref{fig:omc}) even when the ions become effectively unmagnetized at low cyclotron frequency.   As we discuss in \S \ref{sec:interp}, this lack of growth at $\omega_{cp} \lesssim \Omega$ for $\vec{\Omega}$ and $\vec{B}$ anti-parallel is likely an artifact of restricting our numerical solutions to $\vec{k} = k_z \ez$ only.   

Taken together, Figures \ref{fig:DR}-\ref{fig:DR-anti} demonstrate that the physics of kinetic ions and electrons dramatically change the properties of instabilities driven by differential rotation in high $\beta$ low collisionality plasmas relative to that predicted by the MHD or Hall MHD theory of the MRI.   Ion-driven instabilities grow on large scales $k_z v_a \ll \Omega$ where magnetic tension is irrelevant.  In addition, there is a new kinetic electron instability with growth rates $\gg \Omega$, far exceeding previously known instabilities driven by differential rotation.

\subsection{Analytic Theory for Kinetic Ions}
\label{sec:analytics-i}

In this section we analytically derive an approximate dispersion relation for the large-scale ($k v_a \ll \Omega$) unstable mode shown in  Figures \ref{fig:DR}-\ref{fig:DR-anti}, and provide a physical interpretation of the results.   On these scales magnetic tension is negligible so the physics of the instability is quite different from that of the more familiar ideal MHD theory of the MRI.   

As is evident from Figures~\ref{fig:DR}-\ref{fig:DR-anti}, the large scale modes are not influenced significantly by electron dynamics. We may thus treat the electrons as a cold, massless, charge-neutralizing fluid (the Vlasov-fluid approximation, see Freidberg 1972).  For the sake of brevity we shall also simplify notation in this section so that that species dependent quantities without a subscript are proton quantities.

For  parallel modes with $b_z = \pm 1$ and $\vec{k} = k_z \ez$, the linear response tensor $\mathbf{\Lambda}$ required to calculate the dispersion relation (eq. \ref{eq:shearing-sheet-dispersion-relation}) is given by equations HQ71 and HQ72.   The arguments of the plasma dispersion function are
\be
\zeta_{\pm} = \frac{\omega \pm \omega_g}{k v_t},
\label{eq:zeta}
\ee
where $k=|k_z|$, 
\be
\omega_g^2 = (1-\Delta) S_z^2 \quad\textrm{}\quad S_z = \omega_c b_z + 2\Omega
\label{eq:omega_g}
\ee
and $\omg > 0$ is the gyration frequency that characterizes the linear orbits of particles due to the combined effects of the rotation and the background magnetic field.
The low-$k$ peak in the dispersion relation in Figure \ref{fig:DR} due to kinetic ions can be derived analytically by assuming the ordering 
\be
\beta\gg\omega_c/\Omega\gg 1
\label{eq:ion-ordering}
\ee
and looking for modes with
\be
\omega/\Omega \sim kv_t/(\omega_c\Omega)^{1/2} \sim 1.
\label{eq:mode-ordering}
\ee
The first inequality in equation \ref{eq:ion-ordering} will always be satisfied for sufficiently weak magnetic fields. The second equality says that we are roughly in the guiding center limit (though in practice the analytic theory is a reasonable approximation even for $\omega_c \sim \Omega$). In the following we will refer to equations \ref{eq:ion-ordering} and \ref{eq:mode-ordering} together as the \emph{gyroviscous ordering}. Note that according to this ordering we have
\begin{equation}
  v_a/\Omega \ll k^{-1} \ll v_t/\Omega.
\end{equation}
The modes under consideration are thus large scale compared to the MRI scale $v_a/\Omega$ but small scale compared to the thermal scale height $v_t/\Omega$ of the disk.

Physically, the ratio $k v_t/(\omega_c\Omega)^{1/2}$ is the ratio of the rate of angular momentum redistribution by viscous stresses (off-diagonal components of the pressure tensor) to the rotation rate.   This is most easily seen in the fluid theory of \citet{Braginskii1965}, in which the anisotropic momentum transport in a collisional, magnetized plasma contains a contribution along the magnetic field (that depends on the collision frequency but is independent of the cyclotron frequency) and a cross-field component (the gyroviscous stress) that depends on the cyclotron frequency but is independent of the collision frequency (see \S \ref{sec:gyroviscous} below for more details).    Thus the ordering in equation \ref{eq:mode-ordering} implies that gyroviscous transport of angular momentum is dynamically important. Magnetic tension on the other hand is negligible. Indeed, the gyroviscous ordering implies
\be
\frac{kv_a}{\Omega}
\sim {\left(\frac{\omega_c}{\beta\Omega}\right)}^{1/2} \ll 1.
\label{eq:no-tension}
\ee
Magnetic tension thus has no effect on the dynamics of the modes under consideration (which have $\omega\sim\Omega$).

Given the gyroviscous ordering, it also follows that the arguments of the plasma dispersion function
\be
\zeta_\pm\sim {\left(\frac{\omega_c}{\Omega}\right)}^{1/2} \gg 1.
\label{eq:large-zeta}
\ee
We may thus employ the large argument expansion of the plasma dispersion function: $W(\zeta)\simeq-\zeta^{-2}$. 

For parallel modes with $b_z = \pm 1$ and $\vec{k} = k_z \ez$, the unstable branch of the dispersion relation is obtained from the perpendicular (with respect to $\ez$) dispersion tensor $\mathbf{D}_\perp$ in equation \ref{eq:shearing-sheet-dispersion-relation}. We wish to find the leading order incarnation of this tensor in the gyroviscous ordering. In order to do so, we introducing the ordering parameter $\epsilon\ll1$ such that $\beta\sim1/\epsilon^2$ and $\omega_c/\Omega\sim1/\epsilon$. It is not difficult to verify that $kv_a/\Omega\sim\epsilon^{1/2}\ll1$ and $\zeta_\pm\sim1/\epsilon^{1/2}\gg1$, consistent with equations \ref{eq:no-tension} and \ref{eq:large-zeta}.

The matrix representation of $\mathbf{D}_\perp$ in the basis $(\ex,\ey)$ is given in equation HQ73. Letting the gyroviscous ordering parameter $\epsilon\to0$ yields
\be
\mathbf{D}_\perp =
\left[\begin{matrix}
  -\omega^2 &
  \displaystyle
  i\omega \left(2\Omega - \frac{k^2 v_t^2}{\omega_c b_z}\right) \\
  \displaystyle
  -i\omega \left(2\Omega - \frac{k^2 v_t^2}{\omega_c b_z}\right) &
  -\omega^2 - 2q\Omega^2
\end{matrix}\right].
\label{eq:D-perp}
\ee
In deriving this expression we have used the ``circular components'' of the response tensor (see HQ71 \& HQ72), which are approximately given by
\be
\Lambda_\pm =
\pm\frac{\omega}{\omega_c}
-\frac{\omega^2}{\omega_c^2}
\mp\frac{\omega}{\omega_c^3} \left[\left(2 - \frac{q}{2}\right)\Omega\omega_c b_z - k^2 v_t^2\right]
+ O(\epsilon^3)
\label{eq:Lambda-pm}
\ee
and
\be
\omega_g/S_z = b_z - q\Omega/(2\omega_c) + O(\epsilon^2).
\label{eq:omg-Sz}
\ee Substituting equations \ref{eq:Lambda-pm} and \ref{eq:omg-Sz} into
equations HQ73 and HQ74 and letting $\epsilon\to0$ also leads directly to
equation \ref{eq:D-perp}. Setting the determinant of equation \ref{eq:D-perp} to zero
yields the dispersion relation \be 
\omega^2\left[ \omega^2 - \kappa^2 + \alpha \Omega^2 \left(4 - \alpha \right)
\right] = 0,
\label{eq:DR-ion}
\ee
where 
\be
\alpha = \frac{k^2 v_t^2}{\Omega \omega_c b_z}
\label{eq:alpha}
\ee
and $\kappa^2=2(2-q)\Omega^2$ is the square of the epicyclic frequency. We stress again that this dispersion relation contains no trace of magnetic tension.

It is straightforward to show from equation \ref{eq:DR-ion} that for $b_z > 0$ the maximum growth rate
\be
\gamma_{\rm max} = \left(-\frac{d\Omega^2}{d\ln r}\right)^{1/2}
= \sqrt{2q}\,\Omega
\label{eq:wmax-ion}
\ee
occurs at
\be
 k_{\rm max} v_t = (2 \Omega \omega_c b_z)^{1/2}.
\label{eq:kmax-ion}
\ee
The maximum growth rate and wavenumber predicted by equations \ref{eq:wmax-ion} \& \ref{eq:kmax-ion} are  in very good agreement with the numerical solutions show in Figures \ref{fig:DR}-\ref{fig:beta} (see, in particular, the dependence on $\omega_c$ in Fig. \ref{fig:omc}).   {  One can also readily show that growth in the gyroviscous limit only occurs for 
\be
2 - \sqrt{2q} < \frac{k^2 v_t^2}{\Omega \omega_c} < 2 + \sqrt{2q}.
\label{eq:krange-ion}
\ee
For $q = 3/2$, i.e., $\kappa = \Omega$, this corresponds to growth for $k v_t/\sqrt{\Omega \omega_c} \,\in \, [0.52,1.9]$.   This modest range in $k$ for the ion-scale gyroviscous mode explains the unusual shape of the dispersion relation curve for $\omega_c/\Omega = 100$ in Figure \ref{fig:DR} (see also Fig. \ref{fig:DR-gyro} discussed below).}

Equation \ref{eq:kmax-ion} implies that the wavelength of the fastest growing mode in the very weak field limit is smaller than the thermal scale height of the plasma $\sim v_t/\Omega$ by only a factor of $\sim (\Omega/\omega_c)^{1/2}$.   This is  contrary to the intuition from the MHD theory of the MRI, where tension requires that growth is restricted to very small scales $\sim v_a/\Omega$ when the magnetic field is weak.   

The maximum growth rate derived here (eq. \ref{eq:wmax-ion}) is identical to that derived by \citet{Quataert2002} and \citet{Balbus2004} in the case of very different physics:   $B_y \ne 0$ and guiding center kinetic theory or anisotropic viscosity along magnetic field lines, respectively.   In both of the latter approximations the ion Larmor motion is averaged out, in contrast to the analysis here which requires finite  cyclotron frequency.   We explain this connection in more detail in \S \ref{sec:ion-interp}.

For $b_z = -1$, i.e., $\vec{\Omega}$ and $\vec{B}$ anti-aligned, eq. \ref{eq:DR-ion} predicts that there are no unstable modes.   This is consistent with the significant difference between the $b_z = 1$ and $b_z = -1$ dispersion relations shows in Figures \ref{fig:DR} \& \ref{fig:DR-anti}.   In particular, in our numerical solutions, the maximum growth rate of $\gamma_{\rm max} = \left(-d\Omega^2/d\ln r\right)^{1/2}$ occurs only for $b_z = 1$, consistent with the analytic dispersion relation.   In the numerical solutions, there is growth at long wavelengths for $b_z = -1$ but it is  slow for $\omega_{c} \gg \Omega$ and is not well described by the ordering used in deriving our analytic approximations  (eq. \ref{eq:ion-ordering} \& \ref{eq:mode-ordering}).   

\subsubsection{Comparison to Fluid Theory with  Gyroviscous Stress}
\label{sec:gyroviscous}

In a magnetized collisional plasma with collision frequency $\nu_i \ll \omega_c$, there are three conceptually distinct contributions to the momentum transport \citep{Braginskii1965}:   (1)  transport of momentum along magnetic field lines, which is equivalent to the field-free transport and thus depends on $\nu_i$ but not $\omega_c$, (2)  cross-field transport which is smaller than the field-aligned transport by a factor of $\sim (\omega_c/\nu_i)^2 \gg 1$, and (3) cross-field transport which is {\em independent} of the collision frequency and is thus suppressed relative to the field-aligned transport by a factor of $\sim (\omega_c/\nu_i)$.   The latter is an example of the gyroviscous stress, which is in general the component of the stress in a magnetized plasma that is perpendicular to the magnetic field and independent of the collision frequency \citep{Ramos2005}.  Physically, this momentum transport arises because of spatial variations in drifts across the Larmor orbits of particles \citep{Kaufman1960}.  For example, in the presence of a background shear, the mean $E \times B$ velocity in a plasma varies spatially across a Larmor orbit, generating a net momentum flux.   The resulting cross-field gyroviscous stress is the leading order finite Larmor radius contribution to the momentum transport.

\citet{Ferraro2007}  derived the dispersion relation for the MRI in a magnetized, collisional plasma accounting for finite Larmor radius effects via the inclusion of gyroviscosity.   He correctly pointed out that at high $\beta$, the dominant correction to the ideal MHD theory of the MRI is not the Hall effect but rather gyroviscosity.   Ferraro also emphasized that gyroviscosity can stabilize short wavelength modes for which magnetic tension is destabilizing in MHD.   He did not, however,  explain the physics of the instabilities that remain in the presence  of gyroviscosity, nor identify the fact that they are physically quite distinct from the MRI.   Given this, and the close connection between our kinetic theory instability calculation and the analogous fluid calculation with Braginskii gyroviscosity,  we find it useful to briefly summarize the fluid theory of the MRI with gyroviscosity.   

In the collisional, magnetized limit, $k v_{t} \ll \nu_i \ll \omega_c$, the ion gyroviscous stress is given by \citep{Braginskii1965}
\begin{equation}
\mathbf{P}^{\rm gv} = \frac{\rho v_{t}^2}{4 \omega_c} \left[\vec{b}\times\!\mathbf{W}\!\cdot\!\Bigl(\mathbf{1} + 3\vec{b}\vec{b}\Bigr)-\Bigl(\mathbf{1} + 3\vec{b}\vec{b}\Bigr)\!\cdot\!\mathbf{W}\!\times\vec{b}\right]
\label{eq:gyrostress}
\end{equation}
where $\mathbf{1}$ is the unit tensor and $W_{ij} = \partial_i u_j + \partial_j u_i - \tfrac{2}{3}\delta_{ij} \nabla\cdot\vec{u}$ is the standard fluid strain tensor.   

It is straightforward to solve for the MHD linear dispersion relation of the shearing sheet including equation \ref{eq:gyrostress} as the only non-ideal MHD term in the equations.   {  We reiterate that neglecting the Hall effect is self-consistent in the limit $\beta \gg 1$.   In addition, parallel viscosity along field lines is unimportant  for the case considered here with {$\vec{k} \parallel \vec{\Omega} \parallel \vec{B}$}.}   For simplicity of presentation we also neglect magnetic tension to be consistent with the analytic approximations to our kinetic theory results derived in \S \ref{sec:analytics-i}.      The resulting dispersion relation in the Boussinesq approximation is
\begin{equation}
\omega^4 - \omega^2 \, \Omega^2 \left(\alpha-2\right) \left(\alpha-\frac{\kappa^2}{2 \Omega^2}\right) + q^2 \Omega^4 \left(\frac{\alpha^2}{4}-\alpha\right)  = 0,
\label{eq:DR-gyroviscous}
\end{equation}
where again $\alpha = (k v_t)^2/(\Omega \omega_{c} b_z)$.   Equation \ref{eq:DR-gyroviscous} is equivalent to \citet{Ferraro2007}'s eq. 10 with his $K \rightarrow 0$.   Equation \ref{eq:DR-gyroviscous} is also the fluid analog of our analytic kinetic theory dispersion relation in the gyroviscous ordering (eq. \ref{eq:DR-ion}).

Equation \ref{eq:DR-gyroviscous} can be solved directly to yield the growing mode predicted by MHD with Braginskii gyroviscosity.   {  The unstable root of the dispersion relation corresponds to the negative branch, i.e.
\begin{align}
 \omega^2 = & \ \frac{1}{2} \, \Omega^2 \, \bigg[(\alpha-2)(\alpha-2+q) 
 \nonumber \\ & - \, \left[(\alpha-2)^2 (\alpha -2 + q)^2 + \alpha \, q^2 (4 - \alpha)\right]^{1/2}\bigg].
 \label{eq:gv-soln}
\end{align}
Equation \ref{eq:gv-soln} can be solved for the wavevector and growth rate of the fastest growing mode but the solution is sufficiently unwieldy as to not provide much insight (this is because the resulting equation is a quartic in $k_{\rm max}^2$).   For $q = 3/2$ and $b_z > 0$, the result is $k_{\rm max} v_t  \simeq 1.2 \, \sqrt{\Omega \omega_c}$ with a corresponding growth rate of $\gamma_{\rm max}/\Omega \simeq  1.31$.   Note that the growth rate of the fastest growing mode in Braginskii theory is somewhat less than the corresponding kinetic theory result in equation \ref{eq:wmax-ion}.   The instability criterion with Braginskii gyroviscosity can be readily determined from the dispersion relation in equation \ref{eq:DR-gyroviscous} evaluated for $\omega = 0$.   This yields that the condition for instability is \citep{Ferraro2007} $0 < \alpha < 4$, i.e.\ $b_z > 0$ and $k v_t < 2 (\Omega \omega_c)^{1/2}$. This implies that growth is restricted to long wavelengths where $k v_A \ll \Omega$, i.e., tension is negligible (so long as $\beta \gg \omega_{cp}/\Omega$).   The solutions of the fluid dispersion relation thus have some similarity to the kinetic dispersion relation, with rapid growth requiring $b_z > 0$ and relatively long wavelengths, $k v_t \lesssim (\Omega \omega_c)^{1/2}$.   However, the fluid and kinetic theory results differ in that in fluid theory there is growth even for $k \rightarrow 0$, while this is not true in kinetic theory (see eq. \ref{eq:krange-ion}).}

\begin{figure}
\centering\includegraphics{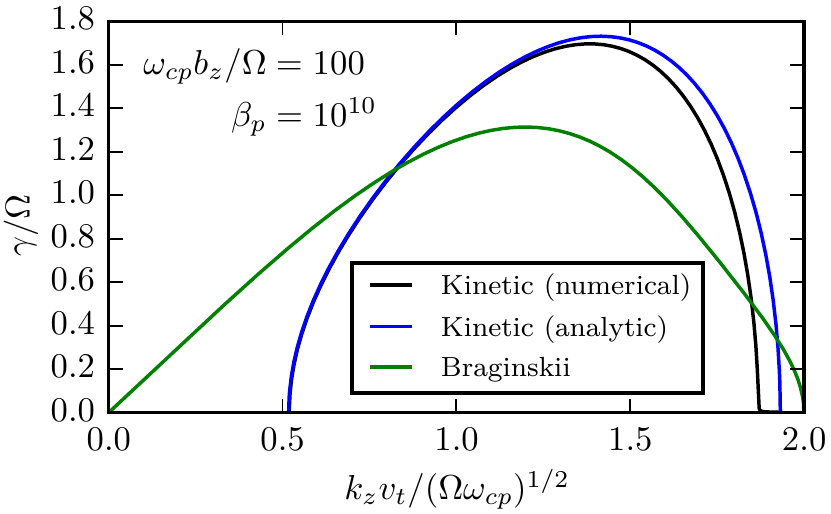}
\caption{Growth rates for ion-driven instabilities in the limit $\beta \gg \omega_{cp}/\Omega \gg 1$.    We compare the full kinetic theory solution for cold, massless electrons, $\beta_p = 10^{10}$, and $\omega_{cp}/\Omega = 100$ (black line) with our analytic kinetic theory dispersion relation (blue line; eq. \ref{eq:DR-ion}) and the analogous collisional fluid theory dispersion relation including the Braginskii gyroviscous stress (green line; eq. \ref{eq:DR-gyroviscous}; \citealt{Ferraro2007}).   The modest quantitative differences between the kinetic and Braginskii results are due to different models for the cross-field viscous transport in these two regimes (\S \ref{sec:ion-interp}).   Note that the analytic solutions are independent of $\beta$ and $\omega_{cp}/\Omega$ so long as
$\beta \gg \omega_{cp}/\Omega \gg 1$ and $k_z$ is normalized as on the x-axis.\label{fig:DR-gyro}}
\end{figure}

Figure \ref{fig:DR-gyro} compares our full kinetic theory numerical dispersion relation (for cold, massless electrons, $b_z = 1$, $\omega_c/\Omega = 100$, and $\beta_p = 10^{10}$) with the kinetic theory analytic approximation (eq. \ref{eq:DR-ion}) and the dispersion relation in MHD with Braginskii gyroviscosity (eq. \ref{eq:gv-soln}).   The kinetic theory analytic approximation is in excellent agreement with the full numerical solution.   The Braginskii gyroviscous model predicts growth over a broader range of wavelengths and with a somewhat smaller peak growth rate.   As we shall now discuss, the difference between the fluid and kinetic results lies in the different form of the viscous stress  in kinetic theory and collisional Braginskii theory.   

\subsubsection{Physical Interpretation of the Ion Instability}
\label{sec:ion-interp}

In MHD, the MRI is driven unstable by the redistribution of angular momentum by magnetic tension.   By contrast,
the numerical and analytic solutions described in \S \ref{sec:analytics-i} show that magnetic tension plays no role in the ion-driven modes present for $\beta \gg \omega_c/\Omega \gtrsim 1$.  This ion instability is thus physically quite distinct from the MHD theory of the MRI.

A closer analog of the ion instabilities described here is the guiding center kinetic theory instability of differentially rotating plasmas studied by \citet{Quataert2002} (and its fluid  analog studied by \citealt{Balbus2004}).  In this `kinetic MRI' (or `magneto-viscous' instability) momentum transport by viscosity (i.e., the off-diagonal components of the pressure tensor in a collisionless plasma) is the key to destabilizing the mode.    In the guiding center limit ($\omega_c \rightarrow \infty$),  transport of angular momentum requires an initial $B_y \ne 0$, which is why \citet{Quataert2002} and \citet{Balbus2004} found significant deviations from the MHD theory of the MRI only for $B_y \ne 0$.    For finite $B_y$, a small perturbation to the initial magnetic field enables viscous stresses to remove angular momentum from the plasma, allowing it to fall inwards.  This distorts the magnetic field in such a way as to promote further viscous redistribution of angular momentum, leading to a runaway.  
 
In guiding center theory, the physics of the MRI in kinetic theory is identical to that in MHD for the case $\vec{k} = k_z\ez$ and $\vec{B} = B_z\ez$.   This is because if momentum transport is solely along magnetic field lines, there is no destabilizing viscous redistribution of angular momentum for $\vec{k} = k_z\ez$ and $\vec{B} = B_z\ez$.    Here, however, we have shown that  finite cyclotron frequency effects produce significant differences relative to MHD  even for $\vec{k} = k_z\ez$ and $\vec{B} = B_z\ez$ (so long as $\beta \gg \omega_{cp}/\Omega$).   The physical interpretation is closely related to the kinetic MRI, but with the viscous stress that transports angular momentum due to cross-field transport associated with a finite ion cyclotron frequency.   To see this, it is helpful to explicitly write out the equations of motion for the Lagrangian  displacement.   In the presence of a finite pressure tensor but neglecting magnetic tension, these take the form:
\be
\frac{\partial^2\vec{\xi}}{\partial t^2}
+ 2\vec{\Omega}\times\frac{\partial\vec{\xi}}{\partial t}
- 2q\Omega^2\xi_x\ex + \frac{\nabla\cdot\delta\mathbf{P}}{\rho} = 0.
\label{eq:EOM}
\ee
We show in the Appendix that the linearly perturbed pressure force in kinetic theory in the shearing sheet is
\be
\frac{\nabla\cdot\delta\mathbf{P}}{\rho} =
\frac{k_z^2 v_t^2}{\omega_c}\frac{\partial\vec{\xi}}{\partial t}
\times\vec{b}
\label{eq:dP-kinetic}
\ee
where we have continued to take $\vec{k} = k_z\ez$. 
Note that when it is expressed in terms of the Lagrangian displacement, the viscous force in equation \ref{eq:dP-kinetic} contains no explicit dependence on the rotation rate (only implicitly, through the fact that the Lagrangian displacement itself evolves differently in a rotating medium).     By contrast, the linearly perturbed gyroviscous stress (eq. \ref{eq:gyrostress}) in the magnetized collisional regime is
\be
\frac{\nabla\cdot\delta\mathbf{P}^{\rm gv}}{\rho} =
\frac{k_z^2 v_t^2}{\omega_c}\left(
  \frac{\partial\vec{\xi}}{\partial t} \times\vec{b}
  + \frac{q\Omega}{2}\vec{\xi}_\perp b_z
\right).
\label{eq:dP-gv}
\ee
{  The dispersion relation  accounting for Braginskii's gyroviscous stress (eq. \ref{eq:DR-gyroviscous}) can be derived by combining the equations of motion in a differentially rotating plasma (eq. \ref{eq:EOM}) with the linearly perturbed gyroviscous stress in equation \ref{eq:dP-gv}.    Likewise, the kinetic theory dispersion relation in the gyroviscous ordering (eq. \ref{eq:DR-ion}) can be derived by combining equations \ref{eq:EOM} and \ref{eq:dP-kinetic}.}

The two expressions above for the viscous stress (eqs \ref{eq:dP-kinetic} \& \ref{eq:dP-gv}) agree in the non-rotating limit $q \Omega = 0$ but not in the rotating case.   In general, the exact form of the gyroviscous stress depends on the plasma conditions and is different for collisionless plasmas vs. highly collisional plasmas \citep{Ramos2005}.   Thus it is not surprising that the kinetic theory and Braginskii models give qualitatively similar but quantitatively different results (Fig. \ref{fig:DR-gyro}).  

The above analysis of the cross-field momentum transport leads to a simple interpretation of the kinetic theory numerical and analytical results.  In particular, for linear perturbations with $k = k_{max}$ given by equation \ref{eq:kmax-ion} the azimuthal viscous force in equation \ref{eq:dP-kinetic} exactly cancels the Coriolis force.   The solution in this case is pure radial motion with $\partial^2 \xi_x/\partial t^2 = 2 q \Omega^2 \xi_x$, which yields the maximum growth rate in equation \ref{eq:wmax-ion}.   This basic physics - pure radial motion due to efficient viscous redistribution of angular momentum - is  similar to that identified by \citet{Quataert2002} and \citet{Balbus2004} in the guiding center limit.    In  guiding center theory, viscous redistribution of angular momentum yields modes with growth rates $\sim \sqrt{2q} \Omega$ over a wide range of $k$, provided  that there is an azimuthal component of the background magnetic field, that magnetic tension is negligible, and that the timescale for viscous stresses to redistribute angular momentum is short compared to the rotation period.  The existence of rapid growth over a wide range of k is a consequence of the fact that in guiding center theory momentum transport is only along field lines and there is no stabilizing radial viscous force for $\vec{k} = k_z \ez$.   By contrast, in the gyroviscous limit $\beta \gg \omega_c/\Omega \gtrsim 1$ there is in general both a stabilizing radial force and a destabilizing azimuthal force (eqs. \ref{eq:dP-kinetic} \& \ref{eq:dP-gv}).   This restricts rapid growth to a modest range of $k$ where the viscous and Coriolis forces are comparable in magnitude (Fig. \ref{fig:DR-gyro}).

\subsection{Analytic Theory for Kinetic Electrons}
\label{sec:analytics-e}

The numerical solutions in Figures \ref{fig:DR}-\ref{fig:DR-anti} show that in the weak magnetic field limit kinetic electrons produce a new unstable mode that can have a growth rate $\gg \Omega$.   This does not have any analog in the previous literature on the MRI.    We now show analytically that this instability is produced by electrons tapping into the free energy of the temperature anisotropy present in a differentially rotating plasma.

Taking $\omega_g \sim \omega_c$ and $k v_a \sim \Omega$, the electron-driven modes in Figures \ref{fig:DR}-\ref{fig:DR-anti} correspond to the argument of the plasma dispersion function (eq. \ref{eq:zeta}) being
 $\zeta_\pm \sim (|\omega_{ce}|/\Omega) \beta^{-1/2} (m_e/m_p)^{1/2}$ for the electrons.   We thus see that for any finite value of $|\omega_{ce}|/\Omega$, the approximation $\zeta_\pm \ll 1$ will be valid for sufficiently large $\beta$.   We make this approximation in what follows and confirm its domain of validity after the fact (see eq. \ref{eq:electron}).   In addition, we utilize the fact that for a proton-electron plasma with $m_p \gg m_e$,  the proton contribution to the total conductivity tensor (eq. \ref{eq:conductivity-sheet}) is negligible for the modes of interest, as can also be checked after the fact.

As in \S \ref{sec:analytics-i}, we use equations HQ71 and HQ72 for the linear response tensor $\mathbf{\Lambda}$ appropriate for parallel modes with $b_z = \pm 1$ and $\vec{k} = k_z \ez$.  For $\zeta_\pm \ll 1$, the plasma response function can be expanded as $W(\zeta) = 1 + i \sqrt{\pi/2} \, \zeta$.   In this case, the linear response tensor simplifies greatly and becomes diagonal.  The components transverse to the magnetic field are given by
\be
\left[\mathbf{Q} \cdot \mathbf{\Lambda} \cdot \mathbf{Q}\right]_\perp = - \frac{i\omega \sqrt{\pi/2}}{kv_t} \Big[ \ex \ex + \left(1-\Delta \right) \ey \ey \Big].
\label{eq:response_par}
\ee
With equation \ref{eq:response_par}, it follows directly from equation \ref{eq:shearing-sheet-dispersion-relation} that the dispersion tensor itself is diagonal.  As a result, the dispersion relation factors into three simple contributions, one for each diagonal component of the dispersion tensor.  The unstable branch is the $yy$ component, which yields the dispersion relation:
\begin{equation}
\omega = \frac{-ikv_{te}}{\sqrt{\pi/2}(1-\Delta_e)}\left[\frac{k^2 v_a^2}{\omega_{cp}^2} \frac{m_e}{m_p} + \frac{2 \Delta_p \Delta_e}{q}\right]  
\label{eq:DR-par}
\end{equation}
\begin{align}
 \ \ \ = \ & \frac{-ikv_{te}}{\sqrt{\pi/2}}  \bigg[\frac{k^2 v_a^2}{\omega_{cp}^2} \frac{m_e}{m_p}\frac{1}{1-\Delta_e}   \nonumber \\
     & + \ \frac{2 q \Omega^2}{(\omega_{cp} b_z + 2\Omega)(\omega_{ce} b_z + \kappa^2/2 \Omega)}\bigg].   \nonumber 
\end{align}
{  The first term in [ ] in the top line of equation \ref{eq:DR-par} can also be written as $k^2 \ell_e^2$ where $\ell_e$ is the electron inertial length.  This will be useful below.    }

In equation \ref{eq:DR-par}, we have  written the dispersion relation both in terms of the electron and proton tidal anisotropies $\Delta_e$ and $\Delta_p$, respectively, and explicitly in terms of the cyclotron and orbital frequencies using equation \ref{eq:tidal-anisotropy}.    
Note that we restrict ourselves to $\Delta < 1$ in order for the background circular orbits to be stable (eq. \ref{eq:omega_g}).   Otherwise the entire shearing sheet formalism is suspect.    Thus the first term in equation \ref{eq:DR-par} has Im$(\omega) < 0$, which is stabilizing, i.e., damping.   The destabilizing term is the second term in [ ].   The condition for instability is thus that $\Delta_p \Delta_e/q < 0$, i.e., that
\be
\label{eq:instability}
\frac{2q\Omega^2}{(\omega_{cp} b_z + 2 \Omega)(\omega_{ce} b_z + 2 \Omega)} < 0 \ \ \ \ {\rm (Instability)}
\ee
We focus  on the case $q > 0$, which is appropriate for a typical astrophysical rotation law.  
The condition for instability depends on the ratio of the disk rotation frequency to the cyclotron frequencies of the particles:

\noindent{\bf 1.}   $|\omega_{ce}| \lesssim \Omega$:   there is no instability for $q > 0$.   This suggests that  an unmagnetized plasma ($|\omega_{cs}|/\Omega \rightarrow 0$) is linearly stable.   As we discuss in \S \ref{sec:discussion}, however, the unmagnetized shearing sheet is in fact linearly {\em unstable} but to non-axisymmetric modes not considered in equations \ref{eq:DR-par} and \ref{eq:instability}.

\noindent{\bf 2.}    $\omega_{cp} \gtrsim \Omega$:  because $\omega_{ce}$ and $\omega_{cp}$ have opposite signs, equation \ref{eq:instability} is always satisfied and there are unstable modes for both $b_z = \pm 1$, i.e., for $\vec{\Omega}$ and $\vec{B}$ parallel and anti-parallel.

\noindent{\bf 3.}   $\omega_{cp} \lesssim \Omega \lesssim |\omega_{ce}|$:   there are only unstable axisymmetric modes for $b_z > 0$ (i.e., $\vec{\Omega}\cdot\vec{B} > 0$) given our sign convention that $\omega_{ce} < 0$.   As we discuss in \S \ref{sec:interp}, however, there are very likely unstable modes for $b_z < 0$, but only for non-axisymmetric wavevectors not considered in eqs. \ref{eq:DR-par} and \ref{eq:instability}.

{  Equation \ref{eq:DR-par} predicts growth for 
\begin{align}
\label{eq:e-krange}
\frac{k v_a}{\Omega} < \sqrt{2q} \ \ \ \ \ \ \ \ \ \ (\omega_{cp} \gtrsim \Omega) \nonumber \\
\frac{k v_a}{\Omega} < \left(q \, \frac{\omega_{cp} b_z}{\Omega}\right)^{1/2} \ \ \ \ (\omega_{cp} \lesssim \Omega \lesssim |\omega_{ce}|)
\end{align}
though the numerical solutions in Figure \ref{fig:DR} show that the instability ceases to be driven primarily by the electrons at sufficiently low $k v_a/\Omega$.   The fastest growing electron mode associated with equation \ref{eq:DR-par} has}
\begin{align}
\label{eq:kmax}
\frac{k_{\rm max} v_a}{\Omega} & = \left(\frac{-2 \Delta_e  \Delta_p}{3 \, q} \frac{m_p}{m_e} \frac{\omega_{cp}^2}{\Omega^2}\right)^{1/2} \\
& \simeq \left(\frac{2 q}{3}\right)^{1/2} \ \ \ \ \ \ \ \ \ \ (\omega_{cp} \gtrsim \Omega) \nonumber \\
& \simeq  \left(\frac{q}{3} \frac{\omega_{cp} b_z}{\Omega}\right)^{1/2} \ \ \ \ (\omega_{cp} \lesssim \Omega \lesssim |\omega_{ce}|) \nonumber
\end{align}
This can also be written more compactly in terms of the electron intertial length as
\begin{equation}
k_{\rm max} \ell_e = \left(\frac{-2 \Delta_e  \Delta_p}{3 \, q}\right)^{1/2}.
\end{equation}
The results for the wavelength of the fastest growing modes in equation \ref{eq:kmax}
are very similar to (though not identical to) the Hall MHD results for the fastest growing MRI modes \citep{Wardle1999}.   The maximum growth rates are, however, very different.    In particular, the fastest growing modes have
\begin{align}
\label{eq:om-max}
\gamma_{\rm max} & \simeq  0.9 \, \Omega \, \beta_e^{1/2} q^{3/2} \left(\frac{\Omega}{\omega_{cp}}\right)^2 \left(\frac{m_e}{m_p}\right)^{1/2} \  (\omega_{cp} \gtrsim \Omega) \\
 & \simeq 0.3 \, \Omega \, \beta_e^{1/2} q^{3/2} \left(\frac{\Omega}{|\omega_{ce}| b_z}\right)^{1/2} \hspace{0.6cm} (\omega_{cp} \lesssim \Omega \lesssim |\omega_{ce}|) \nonumber
\end{align}
Equation \ref{eq:om-max} shows analytically that in the weak field limit of high $\beta$ and finite $\omega_c/\Omega$, the growth rate can be $\gg \Omega$, in contrast to all existing known limits of the MRI.   Moreover, this rapid growth requires kinetic electrons.    

It is important to reiterate that the analytic results of this section (and, in particular, eq.~\ref{eq:DR-par}) were derived assuming that $\zeta_\pm \ll 1$ (eq. \ref{eq:zeta}) for the electrons.   This constraint requires (for the fastest growing mode)
\begin{equation}
\beta_e \gg \frac{m_e}{m_p} \left(\frac{\omega_{ce}}{\Omega}\right)^2  \ \ \  \ \ (\omega_{cp} \gtrsim \Omega) \nonumber
\end{equation}
\begin{equation}
\beta_e \gg \frac{|\omega_{ce}|}{\Omega}  \ \ \  \ \ \ \ (\omega_{cp} \lesssim \Omega \lesssim |\omega_{ce}|).
\label{eq:electron}
\end{equation}
Note that since the growth rates increase $\propto \beta_e^{1/2}$ and the analytic solutions are only valid for sufficiently large $\beta_e$,  this implies that the analytic results are valid precisely when the growth rates are $\gg \Omega$, i.e., in the regime of most interest.  

\begin{figure}
\centering\includegraphics{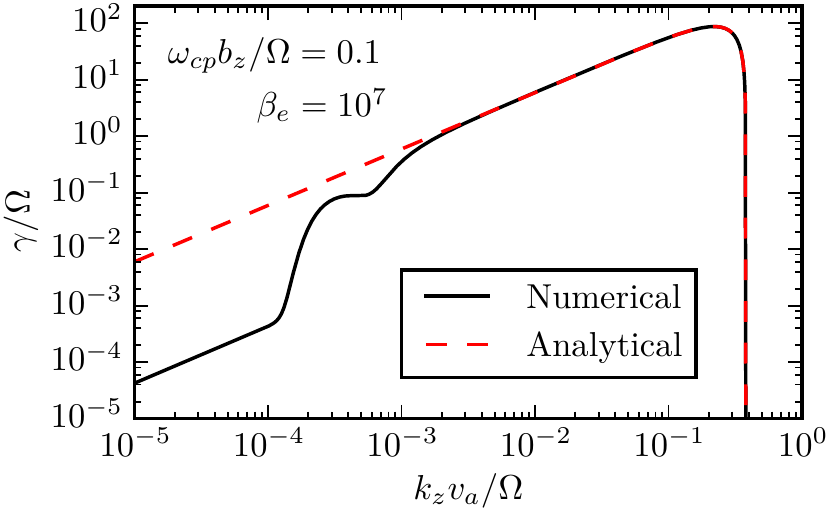}
\caption{Numerical growth rates for the electron-driven instability (solid black line) compared to the analytic dispersion relation in equation \ref{eq:DR-par} (dashed red line).   The agreement is excellent for the fastest growing modes.   At low $k_z v_a/\Omega$, the  expansion of the plasma dispersion function used in the analytic derivation is inapplicable.\label{fig:DR-electron}}
\end{figure}

The analytic results in equations \ref{eq:DR-par}-\ref{eq:om-max} are in good agreement with the numerical solutions of the kinetic dispersion relation discussed in \S \ref{sec:numerics}.   {   We compare the two directly in Figure \ref{fig:DR-electron}.  At sufficiently low $k_ z v_A/\Omega$, the low $\zeta$ expansion of the plasma dispersion function used in the analytic derivation breaks down, but the analytics are an excellent approximation for the fastest growing modes.   More generally, compared to the numerical solutions in  Figures \ref{fig:DR}-\ref{fig:DR-anti}, } the analytic results correctly capture that the fastest growing mode for sufficiently high $\beta_e$ has (1) a growth rate $\gg \Omega$ and $\propto \beta_e^{1/2}$ (2) a growth rate $\propto (\Omega/\omega_{cp})^{1/2}$ for $\omega_{cp} \ll \Omega$ and $b_z = 1$, and (3)  a wavelength of $k v_a/\Omega \sim \min[1,(\omega_{cp}/\Omega)^{1/2}]$.  In addition, the analytics confirm that for $\omega_{cp} \gtrsim \Omega$, the kinetic electron instability is independent of the sign of $b_z$ (compare Figs. \ref{fig:DR} and \ref{fig:DR-anti}).   We find analytically that there are no instabilities associated with kinetic electrons for $b_z = -1$ and $\omega_{cp} \lesssim \Omega$ (eq. \ref{eq:instability}).  This is consistent with our inability to find any growing modes numerically in this regime.    

\subsubsection{Physical Interpretation of the Electron Instability}
\label{sec:interp}

It is straightforward to combine equations \ref{eq:tidal-anisotropy}, \ref{eq:temp-anisotropy}, \& \ref{eq:DR-par} to show that the dispersion relation can be written explicitly in terms of the temperature anisotropy imposed by differential rotation:
\begin{align}
\label{eq:whistler}
& \omega   =   \frac{-ikv_{te}}{\sqrt{\pi/2}} \, \left( \frac{2}{q} \left[\frac{T_{x e}-T_{y e}}{T_{y e}} \right] \left[ \frac{T_{x i}-T_{y i}}{T_{x i}}\right]
+ k^2 \ell_e^2 \, \frac{T_{x e}}{T_{y e}}\right) \nonumber \\
& \simeq \frac{-ikv_{te}}{\sqrt{\pi/2}} \, \left(\frac{T_{x e}-T_{y e}}{T_{y e}} + k^2 \ell_e^2 \right)  \ (\omega_{cp} \lesssim \Omega \lesssim |\omega_{ce}|)
\end{align}
where $\ell_e$ is the electron inertial length.   Equation \ref{eq:whistler} is easiest to interpret when $\omega_{cp} \lesssim \Omega \lesssim |\omega_{ce}|$, in which case the dispersion relation reduces to the approximate equality on the second line.

Because the unstable root here is associated with the $yy$ component of the dispersion tensor, it is also straightforward to show that the polarization of the unstable mode is
\begin{align}
\label{eq:polarization}
& \delta {\tilde E_z} = 0 \ \ \ \  \ \delta {\tilde E_x} =   \frac{q \Omega}{i \omega} \delta {\tilde E_y} \nonumber \\
& \delta \vec{B} =  \frac{- k_z \, \delta {\tilde E_y}}{\omega} \, \ex \nonumber \\
& \delta \vec{J} = \frac{- i k_z^2 \delta {\tilde E_y}}{\omega \mu_0} \, \ey 
\end{align}
Note that equation \ref{eq:polarization} implies that the linearly unstable mode has no Maxwell or Reynolds stress, in contrast to the MRI.    In addition, substituting eqs. \ref{eq:response_par} and \ref{eq:polarization}  
in equation \ref{eq:dP} shows that the linearly perturbed pressure force also vanishes.  

Equations  \ref{eq:whistler} and \ref{eq:polarization} demonstrate that the unstable mode is a circularly polarized electron whistler driven unstable by the background temperature anisotropy (which is itself created by the differential rotation).   The resulting instability is also  closely related to the electron Weibel instability of an unmagnetized plasma \citep{Weibel1959}:   the second line of equation \ref{eq:whistler} is identical to the electron Weibel instability dispersion relation in the limit of a small fractional electron temperature anisotropy.   This result is not particular to the shearing sheet:   in a homogeneous magnetized plasma with a gyrotropic temperature anisotropy $\Delta T_e \lesssim T_e$, the electron Weibel instability persists as an electron whistler instability so long as $|\omega_{ce}| \ll kv_{te}$, which corresponds to $v_a/v_{te} \ll (m_e/m_p)(\Delta T_e/T_e)^{1/2}$ for the fastest growing mode.

Equation \ref{eq:whistler} shows that for $\omega_{cp} \lesssim \Omega \lesssim |\omega_{ce}|$ instability is present only if $T_{x \, e} < T_{y \, e}$.   Given equations  \ref{eq:tidal-anisotropy} \& \ref{eq:temp-anisotropy}, this is equivalent to the constraint $b_z > 0$, i.e., $\vec{\Omega}\cdot\vec{B} > 0$, noted in \S \ref{sec:analytics-e}.   
 In the velocity space instability interpretation provided here, this constraint arises for the following reason:   the Weibel instability in a homogeneous plasma requires that the wavevector have a component along the {\em low} temperature direction.    For our equilibrium shearing sheet model with $\omega_{cp} \lesssim \Omega \lesssim |\omega_{ce}|$ and $\vec{\Omega}\cdot\vec{B} > 0$, $T_x = T_z < T_y$ while for  $\vec{\Omega}\cdot\vec{B} < 0$, $T_x = T_z > T_y$.    Thus modes with $\vec{k} \parallel \vec{\Omega}$ will only be unstable for $\vec{\Omega}\cdot\vec{B} > 0$, as is indeed the case.    This analysis demonstrates, however, that the case $\vec{\Omega}\cdot\vec{B} < 0$ will also be unstable for $\omega_{cp} \lesssim \Omega \lesssim |\omega_{ce}|$, but probably only to non-axisymmetric modes with $k_y \ne 0$ since those will have a projection of the wavevector along the low temperature direction.  A full non-axisymmetric instability calculation would be quite involved, however, so we leave explicit demonstration of this point to future work.    
 
\section{Planar Shear Flows}
\label{sec:shear} 

{  The results derived in the previous sections  can also be applied to study the stability of non-rotating planar shear flows.      Although a planar shear flow is linearly stable in ideal hydrodynamics or MHD, the inclusion of non-ideal physics such as ambipolar diffusion or the Hall effect can generate linear instability \citep{Kunz2008}.   The importance of these linear instabilities is uncertain given the well-known non-linear hydrodynamic instabilities afflicting planar shear flows.   Nonetheless,  the presence of magnetically-mediated linear instabilities in planar shear flows might in some cases alter the resulting turbulence, transport properties, and/or magnetic field amplification relative to that predicted by non-linear hydrodynamic turbulence.

A  shear flow with an equilibrium velocity $\vec{v} = Sx \ey$ satisfies the identical dispersion relation to that derived in \S \ref{sec:HQ} (eq. \ref{eq:DR-par}) with  $\Omega \rightarrow 0$ but $q \Omega \rightarrow -S$ (and thus finite).  Using this transformation, it is straightforward to assess the stability of a planar shear flow to the instabilities of rotating plasmas highlighted in this paper.    Here we briefly summarize the conclusions drawn from making this transformation, but we defer a detailed study of the stability of planar shear flows in the shearing sheet  to future work.    As in the bulk of this paper, we restrict our analysis to $\vec{B} = B_z \ez$ and $\vec{k} = k_z \ez$.

A planar shear flow has an equilibrium temperature anisotropy set by $\Delta_s = -S/\omega_{c s} b_z$ (see eq. \ref{eq:temp-anisotropy}).   We restrict ourselves to $\Delta_s < 1$ for both electrons and protons, so that the equilibrium orbits are stable (this follows from the fact that $\omega_g^2 < 0$ in eq. \ref{eq:omega_g} for $\Delta > 1$).   In this case, a planar shear flow is stable to the electron temperature anisotropy instability described in \S \ref{sec:analytics-e} (at least for $\vec{B} = B_z \ez$ and $\vec{k} = k_z \ez$).  This follows from equation \ref{eq:DR-par} by setting $\Omega \rightarrow 0$ and $q \Omega \rightarrow -S$.    Physically, the reason is that the perturbed ion current associated with the temperature anisotropy exactly cancels the analogous perturbed current due to the electrons.   Mathematically, this corresponds to the fact that the nominally destabilizing term in the dispersion relation (eq. \ref{eq:DR-par}) is the last term $\propto q \Omega^2$, which vanishes for a shear flow.    Since Weibel instabilities are driven by current bunching, the fact that the ion current shields the electron current for a planar shear flow  leads to linear stability.   

A planar shear flow is also linearly stable to the kinetic theory version of the ion gyroviscous instability described in \S \ref{sec:analytics-i}.   This follows from equation \ref{eq:DR-ion} by setting $\Omega \rightarrow 0$ and $q \Omega \rightarrow -S$.\footnote{This conclusion only applies to modes with $\omega \sim S$ and $k v_t \sim \sqrt{S \omega_c}$, assuming $\beta \gg \omega_c/S \gg 1$, which is the shear flow analog of the ordering used in \S \ref{sec:analytics-i}.}     However, a planar shear flow is linearly unstable in the presence of Braginskii gyroviscosity, i.e., in the magnetized, collisional limit.   Indeed, setting $\Omega \rightarrow 0$ and $q \Omega \rightarrow -S$, equation \ref{eq:DR-gyroviscous} becomes
\begin{equation}
\omega^4 - \omega^2 \,  (f^2 - Sf) + \frac{S^2 f^2}{4}  = 0,
\label{eq:DR-gv-shear}
\end{equation}
where $f = k^2 v_t^2/\omega_c b_z$ and  equation \ref{eq:DR-gv-shear} is derived assuming magnetic tension is negligible, which requires $\beta \gg \omega_c/S$.   The solution to equation \ref{eq:DR-gv-shear} is
\begin{equation}
\omega^2 = \frac{1}{2}\left[f(f-S) \pm \left(f^2(f^2 - 2 S f)\right)^{1/2}\right]
\label{eq:soln-sheargv}
\end{equation}
For $f \gg S$, equation \ref{eq:soln-sheargv} corresponds to dispersive waves with $\omega = k^2 v_t^2/\omega_c$.   It is straightforward to show from equation \ref{eq:soln-sheargv}, that a linear shear flow with $\beta \gg \omega_c/S$ is subject to linear {\em over-stabilities} (i.e., solutions that both  oscillate and exponentiate in time) provided that $b_z > 0$ and $f < 2S$.   The fastest growing mode has
\begin{equation}
k_{\rm max} v_t = \left(S \omega_c b_z\right)^{1/2} \ \ {\rm and} \ \ \omega_{\rm max} = \pm \frac{S}{2} + i \frac{S}{2}.
\label{eq:shear-flow}
\end{equation}
The growth rate of the fastest growing mode for the planar shear flow is thus comparable to that for a differentially rotating flow shown in Figure \ref{fig:DR-gyro}.   The physical interpretation of this ion-driven instability of planar shear flows in the magnetized, collisional (Braginskii) limit is also similar to that described for rotating flows in \S \ref{sec:ion-interp}:   a perturbation to the magnetic field generates a  viscous force that displaces the plasma further from its initial equilibrium position,  enhancing the initial perturbation to the magnetic field and the resulting viscous force.    

\section{Application to Virialized Plasma in High Redshift Halos}
\label{sec:halo}

Here we briefly describe the application of our work to the origin of magnetic fields at high redshift.   This is a complex problem whose full solution is well beyond the scope of this paper.     Here we focus on providing simple estimates of magnetic field amplification in the virialized plasma in the outskirts of high redshift dark matter halos, because these plasma conditions are reasonably well understood and because this plasma is the least likely to be magnetized by other processes.   We assume that a seed magnetic field is already present, generated by, e.g., the Biermann battery (e.g., \citealt{Naoz2013}) or Weibel-like instabilities (e.g., \citealt{Lazar2009}; Spitkovsky \& Quataert, in prep).    

We first show that the halo plasma conditions of interest are relatively collisional.  As a result, we argue that the most important instability is likely to be the collisional version of the ion-driven instability described in \S \ref{sec:analytics-i} \& \S \ref{sec:shear}.  We scale typical estimates to dark matter halos with masses of $10^8 M_8 M_\odot$ at redshift $\sim 10-20$.   In particular, $2-3 \, \sigma$ density fluctuations at $z \sim 10$ lead to collapsed halos with masses $\sim 10^8-10^{10} M_\odot$ (e.g., \citealt{Barkana2001}).  The virial temperature of plasma in such halos is
\be
T_{\rm vir} \simeq 2 \times 10^4  \, M_8^{2/3} \left( \frac{1+z}{10} \right) \, {\rm K}
\label{eq:Tvir}
\ee
while the characteristic plasma density in the outer parts of the halo is roughly 200 times the mean baryonic density of the Universe (independent of halo mass):
\be
n \sim 0.05 \, \left(\frac{1+z}{10}\right)^3 \, {\rm cm^{-3}}.
\label{eq:nhalo}
\ee
These estimates compare well to the temperatures and densities in numerical simulations of the formation of high redshift proto-galaxies (e.g., \citealt{Wise2007}).   Given these plasma parameters, the corresponding proton collision frequency and mean free path are
\be
\nu_{p} \sim 10^{-8}  \, M_8^{-1}  \left(\frac{1+z}{10}\right)^{3/2} \, {\rm s^{-1}}
\label{eq:nupp}
\ee
and
\be
\ell_p \sim 10^{14} \, M_8^{4/3}  \left(\frac{1+z}{10}\right)^{-1}  \, {\rm cm}.
\label{eq:lp}
\ee
The electron mean free path is comparable to the proton mean free path while the electron collision rate is $\sqrt{m_p/m_e}$ times larger.   

The characteristic shear rate in the plasma is set by the dynamical time, which is of order a tenth of the Hubble time, i.e.
\be
S \sim t_{\rm dyn}^{-1} \sim 10^{-15} \left(\frac{1+z}{10}\right)^{-3/2} \, {\rm s^{-1}}
\label{eq:shear}
\ee
The typical rotation rate for inflowing matter in the outskirts of dark matter halos is $\Omega \sim 0.1 S$ (e.g., \citealt{Barkana2001}) but rotation of course becomes more important as matter flows in to smaller radii.   

A comparison of equations \ref{eq:nupp} and \ref{eq:shear} shows that on the rotation/shear timescale the plasma is reasonably collisional.   This implies that the temperature anisotropy cannot reach the full value in equation \ref{eq:temp-anisotropy} (which is valid only in a collisionless plasma) but will instead be limited to $\Delta T/T \sim S/\nu$ \citep{Alex2005}.   As a result, the electron temperature anisotropy driven instability derived here cannot be directly applied to the collisional halo plasma.  We defer to future work an investigation of this instability under  collisional conditions.   We stress, however, that there will be regions of much higher temperature and much lower collisionality in high redshift galaxies; e.g., supernova remnants and black hole and/or neutron star accretion flows will occur soon after the formation of the first stars.   The conditions in such regions are not as well understood but it is very plausible that these regions are critical sites of magnetogenesis because the low collisionality conditions enable a wider range of plasma instabilities to be important.   Indeed, we have shown that under low collisionality conditions, electron instabilities can amplify the magnetic field on a timescale much less than the rotation period by tapping into the temperature anisotropy generated by differential rotation.   

The fact that $\nu_{p} \gg S$ in the halo plasma suggests that the most important instability  is likely to be the collisional version of the gyroviscous instability described in \S \ref{sec:analytics-i}, along with the magneto-viscous instability generated by collisional transport of momentum along magnetic field lines \citep{Balbus2004}.   Both of these instabilities require $\omega_c \gtrsim \nu_p$, which corresponds to $B \gtrsim 10^{-12} \, {\rm G} \, M_8^{-1} ([1+z]/10)^{3/2}$ under the conditions of interest.   For our fiducial parameters and a $10^{-12}$ G field, $\beta \sim 10^{12} \gg \omega_c/S \sim 10^7$, so that the gyroviscous stress is indeed dynamically important.    The results of this paper demonstrate that both rotation and planar shear flows in the halos of high redshift galaxies will be unstable to gyroviscosity-mediated instabilities that will exponentially amplify the magnetic field on a timescale comparable to, or somewhat shorter than, the rotation/shear time.    As the magnetic field grows in strength, eventually $\beta \lesssim \omega_c/S$ and gyroviscosity will cease to be dynamically dominant.   At that point, the unstable mode of interest will become the standard MRI.   The ion-driven instabilities described in this paper thus provide a way of amplifying the magnetic field from an initially small value to the point where the magnetohydrodynamic MRI can take over.   The key stage that we have not addressed is how the magnetic field is amplified to the point where $\omega_c \gtrsim \nu_p$, so that Braginskii's collisional, magnetized theory  applies.   }
 
\section{Discussion}
\label{sec:discussion}

We have studied the linear stability of weakly magnetized differentially rotating plasmas  {  in both collisionless kinetic theory and Braginskii's theory of collisional, magnetized plasmas.}   We have focused in particular on the limit of very weak magnetic fields, for which the ion and/or electron cyclotron frequencies are not much larger than the rotation frequency of the plasma.  
Our motivation is primarily to understand how magnetic fields can get created and/or amplified from very small initial values.   Astrophysically, this is particularly important in the context of understanding at what stage during   structure formation at high redshift do magnetic fields become dynamically significant and need to be included in theoretical and numerical models of star formation, galaxy formation, and massive black hole growth.

For very weak magnetic fields, ideal MHD predicts that the most unstable MRI mode driven by differential rotation has a short wavelength $\sim v_a/\Omega$ that gets smaller for weaker fields.   Hall MHD represents an  extension of this theory to $\omega_{cp} \lesssim \Omega$, i.e, to conditions in which the proton cyclotron frequency can be small compared to the disk rotation frequency.    In Hall MHD the MRI persists as a whistler mode destabilized by magnetic tension \citep{Wardle1999}.   The Hall MHD approximation corresponds, however, to the cold ion, cold massless electron limit of kinetic theory.   Thus it is not a good approximation for  high $\beta$ low-collisionality plasmas, nor can it capture any kinetic electron physics.     {  In particular, momentum transport along and across magnetic field lines is a more important non-ideal MHD effect than Hall currents in the high $\beta$ dilute plasmas of interest in this paper.}

These considerations motivate the linear kinetic theory calculation described in this paper, which  self-consistently incorporates finite electron and ion cyclotron frequencies (and Larmor radii).   Our analysis draws heavily on the formalism developed in \citet{HQ2014} (HQ), who carried out a general linear stability calculation in kinetic theory for local instabilities in differentially rotating plasmas. Their primary  assumptions were charge neutrality and axisymmetry.   In this paper we have further restricted our analysis to the simplest non-trivial problem, in which $\vec{B} \parallel \vec{\Omega} \parallel \vec{k}$ (including both parallel and anti-parallel fields).  

The case of  $\vec{B} \parallel \vec{\Omega} \parallel \vec{k}$  captures  the key physics of the MRI in MHD.   Moreover, in guiding center kinetic theory, in which one averages over the Larmor orbits of ions and electrons, the linear theory of the MRI for $\vec{B} \parallel \vec{\Omega} \parallel \vec{k}$ is identical  to that in MHD \citep{Quataert2002}. We have shown, however, that  the case of kinetic ions and electrons with finite cyclotron frequencies is far more interesting.   The single instability of ideal MHD that is mediated by magnetic tension is replaced by two distinct instabilities, one associated with kinetic ions and one with kinetic electrons.  Each of these instabilities has a different way of tapping into the free energy of differential rotation.   

In kinetic theory, if $\beta \gtrsim \omega_{cp}/\Omega \gtrsim 1$, there is an instability at long wavelengths $k v_{tp}/\Omega \sim (\omega_{cp}/\Omega)^{1/2}$ which has a maximum growth rate of $\gamma_{\rm max} = (-d\Omega^2/d\ln r)^{1/2}$ (Figs. \ref{fig:DR} \& \ref{fig:omc} and eqs. \ref{eq:wmax-ion} \& \ref{eq:kmax-ion}).   This instability is associated with kinetic ions and is indifferent to the electron physics, being present for both fluid and kinetic electron models.   Note that for  $\omega_{cp} \sim \Omega$ the wavelength of the fastest growing mode is comparable to the thermal scale height of the plasma $\sim v_{tp}/\Omega$, i.e., the fastest growth is for the largest scale modes.   This is contrary to the predictions of MHD and Hall MHD, in which tension requires that growth is restricted to very small scales when the magnetic field is weak.  

The maximum growth rate found here for the ion instability in the limit of $\vec{B} \parallel \vec{\Omega} \parallel \vec{k}$ is identical to that derived by \citet{Quataert2002}  for the `kinetic MRI' (and its fluid analog, the magnetoviscous instability; \citealt{Balbus2004}), which required $B_\phi \ne 0$ in guiding center theory.   The physical nature of the ion instability found here is also most similar to that described by \citet{Quataert2002} and \citet{Balbus2004}.  At its heart is the fact that viscous transport of angular momentum can be more efficient than magnetic tension in high $\beta$ low-collisionality plasmas.    The angular momentum redistribution is in turn coupled to the magnetic field geometry (because of the Larmor motion of particles), which is what leads to an instability:  perturbations to the initial magnetic field structure enhance the viscous redistribution of angular momentum, allowing plasma to fall inwards, which drags the field with it, further enhancing the  redistribution of angular momentum.  A runaway ensues.   

In guiding center theory, the transport of angular momentum is only along magnetic field lines so that a finite $B_\phi$ is needed to generate viscous transport in linear theory.   By contrast, in our present analysis the momentum transport is due to cross-field terms associated with a finite ion cyclotron frequency.   This cross-field momentum transport is generically known as the gyroviscous stress \citep{Ramos2005}.  We have shown that  this produces an instability driven by viscous transport of angular momentum even for the case of $\vec{B} \parallel \vec{\Omega} \parallel \vec{k}$.   {  Previously, \citet{Ferraro2007}  highlighted the importance of gyroviscosity for the MRI using \citet{Braginskii1965}'s result for the gyroviscous stress in collisional, magnetized plasmas (see \S \ref{sec:gyroviscous}).  He did not, however,   identify the fact that the instabilities that remain in the presence of gyroviscosity are physically quite distinct from the MRI in ideal MHD.

We have shown that the collisionless kinetic theory and Braginskii  models of cross-field gyroviscous transport both produce an instability driven by viscous transport (not magnetic tension) when $\beta \gg \omega_c/\Omega$.  Growth is on similar spatial scales and with similar growth rates in the fluid and collisionless cases, though the kinetic theory growth rates exceed the fluid growth rates by a modest amount (Fig. \ref{fig:DR-gyro}).}  It is important to reiterate that these relatively large scale ion-driven instabilities exist in both the collisionless and magnetized, collisional \citep{Braginskii1965} regimes (see Fig. \ref{fig:DR-gyro}).   This is significant because it implies that they are likely to be relatively robust and present under a wide range of plasma conditions.

In addition to the long-wavelength instability driven by ion momentum transport, we also find a shorter wavelength instability with $k v_a \sim \Omega$ that is present only for the case of kinetic electrons.   This instability is in many ways more remarkable than that due to kinetic ions because the growth rate is $\propto \beta_e^{1/2}$ and can exceed the rotation rate by many orders of magnitude, particular for modest values of $\omega_{cp}/\Omega$ (Figs. \ref{fig:DR}-\ref{fig:DR-anti}).   This is despite the fact that the free energy source for this electron instability is still differential rotation, in the sense that the growth rate is  $\propto d \Omega/dr$.   

We have shown that this kinetic electron instability is the whistler mode driven unstable by the background temperature anisotropy present in the kinetic equilibrium of a differentially rotating plasma (see \S \ref{sec:analytics-e}).   This temperature anisotropy (the `tidal anisotropy'; eqs. \ref{eq:tidal-anisotropy} \& \ref{eq:temp-anisotropy}) is  required to satisfy the Vlasov equation in the equilibrium state.  It is distinct from the more familiar temperature anisotropy relative to a magnetic field typically considered in homogeneous magnetized plasmas.  The kinetic electron instability with growth rates $\gg \Omega$ is thus  a consequence of a unique feature of the {kinetic equilibrium} of a differentially rotating plasma.   This temperature anisotropy is in fact well-known in the theory of collisionless stellar disks, which exhibit an analogous anisotropy (e.g., \citealt{Shu1969}). 

The whistler instability found here is also closely related to the \citet{Weibel1959} instability of unmagnetized plasmas with a temperature anisotropy.  In particular, in the limit $|\omega_{ce}| \gtrsim \Omega \gtrsim \omega_{cp}$, the growth rate of the whistler instability that we have derived (eq. \ref{eq:whistler})  is the same as that of the electron Weibel instability, provided one uses the electron temperature anisotropy implied by the shearing sheet equilibrium (eqs. \ref{eq:tidal-anisotropy} \& \ref{eq:temp-anisotropy}) in  the Weibel dispersion relation.   

This connection is further highlighted by the fact that an initially unmagnetized differentially rotating plasma itself has a temperature anisotropy - with $T_\phi = T_r/4$ for the case of a point mass potential.   Thus an initially unmagnetized differentially rotating plasma is linearly unstable to the electromagnetic \citet{Weibel1959} instability.  Because the Weibel growth rates are much faster than the rotation rate, the background shear has no significant effect on the dynamics of the unstable modes.  The important function of the shear is only that it creates the equilibrium temperature anisotropy.    The existence of the Weibel instability implies that initially unmagnetized low collisionality rotating flows will spontaneously generate a magnetic field on a timescale $\ll \Omega^{-1}$, with the initial length scale of the growing modes of order the electron skin depth.   In a future paper, we will study the saturation of these instabilities using PIC simulations (Spitkovsky \& Quataert, in prep).

The instabilities described in this paper  driven by the temperature anisotropy  in  differentially rotating plasmas are examples of a broader class of instabilities in which the free energy in shear or differential rotation can be tapped via the temperature anisotropy it induces.   For example, in a system nominally described by MHD, the existence of a velocity shear and a finite collisionality  implies that there is a temperature anisotropy $\Delta T/T \sim S/\nu$ where $S$ is the shear rate in the plasma and $\nu$ is the Coulomb collision rate (e.g., \citealt{Alex2005}).  If $|\Delta T/T| \gtrsim \beta^{-1}$, then the system nominally described by MHD will in fact be unstable to a set of velocity-space instabilities including the firehose and mirror instabilities (and, in some cases, the electron whistler and ion cyclotron instabilities).   The impact of these instabilities on the dynamics of astrophysical plasmas remains an area of active investigation (e.g., \citealt{Kunz2014,Riquelme2014}). 

For the $\vec{\Omega} \parallel \vec{B}$ case that we have focused on in this paper, the temperature anisotropy induced by differential rotation is entirely in the plane {\em perpendicular} to the local magnetic field. For the more general case of $B_\phi \ne 0$, the tidal anisotropy will include an anisotropy with respect to the background magnetic field.   Because the tidal temperature anisotropy is $\Delta T/T \sim \Omega/\omega_{c}$ for $\omega_c \gtrsim \Omega$ (see eq. \ref{eq:tidal-anisotropy}) we expect that for $B_\phi \ne 0$ and $\beta \gtrsim \omega_c/\Omega$, differentially rotating plasmas will be unstable to the firehose and mirror instabilities in addition to the whistler instability highlighted in this paper.    This remains to be explicitly demonstrated in future work.  

{  The derivations in this paper can be readily applied to the stability of planar shear flows in addition to differentially rotating plasmas (\S \ref{sec:shear}).   Utilizing this fact, we have found that planar shear flows are subject to linear overstabilities in the presence of the Braginskii gyroviscosity, i.e., in the magnetized, collisional limit.   The resulting overstabilities have growth rates comparable to the shear rate (eq. \ref{eq:shear-flow}).   This suggests that turbulence with a weak magnetic field in the magnetized, collisional regime is likely to be a far richer physics problem than suggested by standard kinematic dynamo models.

The critical question not addressed by our analysis is the ultimate saturation of the instabilities described here and their impact on astrophysical plasmas.    We suspect that these instabilities are important for the amplification of magnetic fields at high redshift.    In particular, we have demonstrated that the virialized plasma in the halos of high redshift galaxies is unstable to instabilities mediated by the collisional, magnetized gyroviscous stress (\S \ref{sec:halo}).  This is true for both rotating flows and planar shear flows.   These instabilities provide a way of amplifying magnetic fields to the point where the canonical MRI of ideal MHD takes over.   

Several important questions remain to be addressed in future work.  In particular, the equilibrium temperature anisotropy in a collisional plasma is significantly less than in a collisionless plasma.  This will decrease the growth rate of the kinetic electron instability found here under many astrophysical conditions (but the instability is so strong that it may nonetheless remain important).  In future work,  it would also be valuable to study the interplay between instabilities driven by cross-field viscous transport of angular momentum (such as the ion instability studied here) and instabilities driven by field-aligned viscous transport (such as those studied by \citealt{Quataert2002} and \citealt{Balbus2004}).   Our focus on $\vec{k} \parallel \vec{\Omega} \parallel \vec{B}$ in this paper precludes the latter from being important (see \S \ref{sec:ion-interp}).   }

\section*{Acknowledgements}   We thank Steve Balbus, Alex Schekochihin,  Sean Ressler, and Greg Hammett for useful conversations and the referee for a particularly thoughtful and constructive report that improved the paper.   This work was supported in part by NSF grants AST-1333682 and PHY11-25915, Simons Investigator awards from the Simons Foundation (to EQ and AS), the David and Lucile Packard Foundation,
and the Thomas Alison Schneider Chair in Physics at UC Berkeley.
                
\bibliography{references}                 

\begin{appendix}

\onecolumn

\section{Divergence of the Pressure Tensor in the Shearing Sheet}

Let $f$ be the one-particle distribution function. The number density $n$
and fluid velocity $\vec{u}$ are then defined as the zeroth and first moment
of the distribution function:
\begin{equation}
  \label{eq:moments}
  n = \int\!d^3v\,f
  \quad\mathrm{and}\quad
  n\vec{u} = \int\!d^3v\,f\vec{v}.
\end{equation}
Taking the second moment with respect to the peculiar velocity
$\vec{v}-\vec{u}$ yields the pressure tensor
\begin{equation}
  \mathbf{P} = \int\!d^3v\,f(\vec{v} - \vec{u})(\vec{v} - \vec{u})
\end{equation}

The Vlasov equation in the shearing sheet is
\begin{equation}
  \label{eq:vlasov-sheet}
  \pder{f}{t} + \vec{v}\cdot\nabla f + \frac{e}{m}
  (\vec{E} + \vec{v}\times\vec{B})\cdot\pder{f}{\vec{v}}
  - (2\vec{\Omega}\times\vec{v} + \nabla\psi)\cdot\pder{f}{\vec{v}} = 0,
\end{equation}
where the tidal potential
\begin{equation}
  \label{eq:tidal-potential}
  \psi = -q\Omega^2 x^2.
\end{equation}
Taking the first moment of \cref{eq:vlasov-sheet} yields the momentum equation
\begin{equation}
  nm\left(
    \pder{\vec{u}}{t} + \vec{u}\cdot\nabla\vec{u}
    + 2\vec{\Omega}\times\vec{u} + \nabla\psi
  \right) + \nabla\cdot\mathbf{P} =
  en(\vec{E} + \vec{u}\times\vec{B}).
\end{equation}
Written in terms of the relative velocity and electric field, defined by
\begin{equation}
  \label{eq:relative-quantities}
  \tvec{u} = \vec{u} + q\Omega x\ey
  \quad\textrm{and}\quad
  \tvec{E} = \vec{E} - q\Omega x\ey\times\vec{B},
\end{equation}
the momentum equation is given by
\begin{equation}
  \label{eq:momentum-sheet}
  nm\left(
    \pder{\tvec{u}}{t} + \tvec{u}\cdot\nabla\tvec{u}
    + 2\vec{\Omega}\times\tvec{u} - q\Omega\tilde{u}_x\ey
  \right) + \nabla\cdot\mathbf{P} =
  en(\tvec{E} + \tvec{u}\times\vec{B}).
\end{equation}
In linear theory, the relative fluid velocity is related to the electric field
via
\begin{equation}
  -i\omega\delta\tvec{u} =
  (\mathbf{Q}\cdot\mathbf{\Lambda}\cdot\mathbf{Q} + \Delta\ey\ey)
  \cdot\left(\mathbf{1} - \frac{q\Omega}{i\omega}\ex\ey\right)\cdot
  \frac{e}{m}\delta\tvec{E}
\end{equation}
Linearizing \cref{eq:momentum-sheet} and solving for the
$\nabla\cdot\delta\mathbf{P}$, the result may be written as
\begin{equation}
\label{eq:dP}
  \nabla\cdot\delta\mathbf{P} =
  \mathbf{T}\cdot en\delta\tvec{E},
\end{equation}
where the tensor
\begin{equation}
  \mathbf{T} = \mathbf{1} + \frac{1}{i\omega}\left[
    -i\omega\mathbf{1}
    + \left(2\vec{\Omega} + \frac{e}{m}\vec{B}\right)\times\mathbf{1}
    - q\Omega\ey\ex
  \right]\cdot(\mathbf{Q}\cdot\mathbf{\Lambda}\cdot\mathbf{Q} + \Delta\ey\ey)
  \cdot\left(\mathbf{1} - \frac{q\Omega}{i\omega}\ex\ey\right).
\end{equation}
\subsection{Gyroviscous Ordering}
Up until now our analysis has been general and can be applied to determine the perturbed pressure tensor in the shearing sheet for any linear wave or instability.  We now specialize to the case of the gyroviscous ordering discussed in \S \ref{sec:analytics-i} (including taking $\vec{k} = k_z \vec{e_z}$), namely
\begin{equation}
  1/\beta\sim\epsilon^2,\quad
  \Omega/\omc\sim\epsilon\ll 1,\quad
  \omega\sim\Omega,\quad
  k^2 v_t^2\sim\omc\Omega.
\end{equation}
In this limit the tensor $\mathbf{T}$ reduces to
\begin{equation}
  \lim_{\epsilon\to0}\,\omc\mathbf{T} =
  -\frac{k^2 v_t^2}{\omc}
  (\mathbf{1} - \vec{b}\vec{b})\cdot
  \left(\mathbf{1} - \frac{q\Omega}{i\omega}\ex\ey\right)
\end{equation}
so that
\begin{equation}
  \lim_{\epsilon\to0}\,\frac{\nabla \cdot \delta\mathbf{P}}{\rho} =
  -\frac{k^2 v_t^2}{\omc}
  (\mathbf{1} - \vec{b}\vec{b})\cdot
  \left(\mathbf{1} - \frac{q\Omega}{i\omega}\ex\ey\right)
  \cdot \frac{\mathbf{\delta\tvec{E}}}{B}.
\end{equation}
    Using the fact that the Hall effect is negligible in the gyroviscous ordering, we can rewrite $\nabla \cdot \mathbf{\delta P}$ in terms of the perturbed ion velocity using $\delta \tvec{E} = - \delta \tvec{u} \times \vec{B}$.   
Rewriting the perturbed velocity $\delta \tvec{u}$ in terms of the Lagrangian displacement using $\partial \vec{\xi}/\partial t = \delta \tvec{u} - \xi_x q \Omega \ey$ yields equation \ref{eq:dP-kinetic} of the main text.
\end{appendix}

\end{document}